\documentclass[lettersize,journal]{IEEEtran}
\usepackage{amssymb,amsfonts}
\usepackage[cmex10]{amsmath}
\usepackage{algorithmic}
\usepackage{algorithm}
\usepackage{array}
\usepackage{subfigure}
\usepackage{amsmath}

\usepackage[caption=false,font=normalsize,labelfont=sf,textfont=sf]{subfig}
\usepackage{textcomp}
\usepackage{stfloats}
\usepackage{url}
\usepackage{verbatim}
\usepackage{graphicx}
\usepackage{cite}
\usepackage{bbm}
\usepackage{enumerate}
\usepackage{tabularx}
\usepackage{fix-cm}
\usepackage[usenames,dvipsnames]{color}
\newtheorem{thm}{Theorem}[section]
\newtheorem{lemma}{Lemma}[section]

\allowdisplaybreaks
\begin{document}
\title{Analysis and Optimization of Robustness in Multiplex Flow Networks Against \\ Cascading Failures}

\author{Orkun İrsoy,  Osman Yağan,~\IEEEmembership{Senior Member,~IEEE}
\thanks{O. İrsoy and O. Yağan are with the Department of Electrical and Computer Engineering, Carnegie Mellon University, Pittsburgh, PA 15213. \tt{oirsoy@andrew.cmu.edu, oyagan@andrew.cmu.edu}. }
\thanks{This work was supported in part by the Air Force Office of Scientific Research (AFOSR) Grant \# FA9550-22-1-0233.}
\thanks{O. İrsoy gratefully acknowledges the support of the Marija Illic Endowed Fellowship in Electrical \& Computer Engineering at Carnegie Mellon University for the 2024–2025 academic year.}
\thanks{This work has been submitted to the IEEE for possible publication. Copyright may be transferred without notice, after which this version may no longer be accessible.}
}

\markboth{Journal of \LaTeX\ Class Files,~Vol., No., Dec~2024}%
{Shell \MakeLowercase{\textit{et al.}}: A Sample Article Using IEEEtran.cls for IEEE Journals}


\maketitle

\begin{abstract}
Networked systems are susceptible to \textit{cascading failures}, where the failure of an initial set of nodes propagates through the network, often leading to system-wide failures.  
In this work, we propose a \textit{multiplex flow network model} to study robustness against cascading failures triggered by random failures.  
The model is inspired by systems where nodes carry or support multiple types of flows, and failures result in the redistribution of flows within the same layer rather than between layers.  
To represent different types of interdependencies between the layers of the multiplex network, we define two cases of failure conditions: \textit{layer-independent overload} and \textit{layer-influenced overload}.  
We provide recursive equations and their solutions to calculate the \textit{steady-state} fraction of surviving nodes, validate them through a set of simulation experiments, and discuss optimal load-capacity allocation strategies.  
Our results demonstrate that allocating the total free space (i.e., capacity minus initial load) to each layer proportional to the mean effective load in the layer and distributing that free space equally among the nodes within the layer ensures maximum robustness.  
The proposed framework for different failure conditions allows us to analyze the two overload conditions presented and can be extended to explore more complex interdependent relationships.  
\end{abstract}


\section{Introduction}

\IEEEPARstart{N}ETWORKED systems, ranging from power distribution to transportation, are integral to many aspects of modern life. 
These networks rarely operate in isolation; instead, usually there are significant interdependencies between different systems.
For example, transportation networks, which include multiple modes of transport, depend on each other as well as the underlying communication and power networks.
In the event of an attack or random failures in an interdependent system, the failures in one of the networks can cause failures of the dependent nodes in the other networks and vice versa. 
This process may continue in a recursive manner, triggering a cascade of failures that can potentially collapse an entire system, often referred as ``\textit{cascading failures}'' \cite{Buldyrev}. 
Notable examples of this vulnerability are the large-scale blackouts caused by minor failures in power supply networks worldwide \cite{blackout_ex_1,blackout_ex_2,blackout_ex_3}.
Yet, such failures are not unique to power grids; rather, they span a variety of application domains such as supply chain management \cite{supply_chain_ex,TANG201658}, natural disaster relief operations \cite{disaster_ex_1,disaster_ex_2}, transportation networks \cite{Zhang_Cheng_Zhao_Li_Lu_Wang_Xiao_2013}, and economic \cite{economy_ex} and social-ecological systems \cite{social_ecological_ex,socio-technical-ex}.

Given its relevance across numerous domains, the robustness of interdependent networks against cascading failures has been an active research field, particularly after the seminal paper of Buldyrev et al. \cite{Buldyrev}. 
However, existing works on cascading failures in interdependent networks usually focus on percolation-based models \cite{Buldyrev, percolation_1, percolation_2, percolation_3, percolation_4, percolation_5}. 
In this model a node can function only if it belongs to the largest connected (i.e., {\em giant}) component of its own network; nodes that lose their connection to this giant core are considered failed. 
Such models are particularly suitable for studying computer/communication networks where connectivity is the fundamental focus of the robustness analysis.
Many real-world systems, however, are tasked with transporting physical commodities or flows; e.g., power networks, transportation networks, supply-chain networks, financial networks, etc. 
In such flow networks, failure of nodes often lead to {\em redistribution} of their load to the functional nodes, potentially overloading and failing them. 
For example, failure of a node in a transportation network would distribute the current passenger load to surviving nodes, potentially causing congestion. 
Similarly, in a financial network, the bankruptcy of a company would impose a financial burden on related businesses, potentially triggering a propagation of failures across the network.
Thus, the dynamics of failures in such cases is governed primarily by flow redistribution rather than the structural changes in the network.

Robustness of flow networks has recently received attention, with the focus being primarily on single networks \cite{flow_1,flow_2,flow_3,flow_4,single_flow_optimizing,Chen_Hu_Meng_Yu_2024,alpha1,alpha2} or multiple {\em interconnected} networks where flow can be redistributed from one network to another \cite{multi_flow_1,scala,zhang_two_flow_redistribution,Kumar_Kumari_Bala_2021,pei_et.al.,hong_suppressing_2016,Wang_Jin_Zhao_2021}.
However, the assumption of load transfer across layers does not account for scenarios where redistribution is confined within individual layers, even though the failure condition of a node depends intricately on its loads across {\em all} layers.  
This distinction highlights a gap in the literature: the study of multiplex flow networks, where nodes support \textit{multiple} distinct commodities or functionalities, each transmitted through a \textit{separate} network layer, and where the \textit{failure condition} of a node is intricately tied to its total load across all layers.
Such systems encompass a wide range of real-life examples, including  management systems where each working unit is assigned multiple tasks, cloud computing systems where individual computing units support different functionalities such as CPU, GPU, and memory, or financial networks where institutions employ different financial instruments; see Section \ref{sec:model} for a comprehensive list of applications.
Here, we aim to fill this gap in the literature by introducing a new ``\textit{multiplex flow network model}'' and comprehensively studying its robustness against cascading failures.

Despite the relevance of such systems, research in this area remains limited.  
Existing studies either focus on specific strategies for coupling nodes between dual layers~\cite{Zhou_Elmokashfi_2017,Ma_Xin_2024}, using topological measures as initial loads, or extend specific models to multiplex settings, such as the \textit{sandpile model}, where cascades are initiated by a random increase in loads~\cite{Lee_Goh_Kim_2012}.  
In contrast, we propose a \textit{multiplex flow network model} with general distributions for initial load-capacity values, accounting for a broader set of systems while providing analytical results for final system size and optimality against cascading failures initiated by random attacks.  
Moreover, the proposed model incorporates more complex overload conditions, capturing intricate dependencies between layers while allowing further extensions for analyzing alternative failure mechanisms.  

The proposed \textit{multiplex flow network model} consists of nodes that operate across multiple layers, with each layer representing a distinct functionality.  
Each node is assigned an initial load and capacity in every layer, and upon failure, the node’s loads are redistributed within the same layer to the remaining functioning nodes. 
These redistributed loads may overload the surviving nodes, potentially triggering further failures depending on the failure condition.  
In our analysis, we consider that if a node fails in one layer (i.e., if its load exceeds its capacity), it also fails in all other layers.  
Regarding failure conditions, we consider two cases:  
i) \textit{Layer-independent overload}, where the overload condition depends only on the capacity and load within the same layer, and  
ii) \textit{Layer-influenced overload}, where loads from other layers also consume a portion of the capacity in the considered layer.  

We evaluate the robustness of the model against cascading failures initiated by random failures/attacks.  
More specifically, the initial attack randomly fails a $p$-fraction of the nodes, causing their loads to be redistributed to the remaining nodes.   Robustness is quantified by the {\em final system size}, i.e., the \textit{steady-state} fraction of surviving nodes. We also consider the \textit{critical attack size} $p^*$ that represents the maximum attack size that the system can sustain with a positive final size; i.e., any initial attack size  greater than  $p^*$  leads to {\em complete} failure of the system. 
Our main contributions include: i) deriving recursive equations for the fraction of surviving nodes at each failure round as a function of the attack size and the distribution of initial loads and capacities (using a mean-field approach); ii) obtaining {\em fixed-point} solutions of the recursive equations and calculating the final system size as a function of all parameters; and iii) deriving the optimal load-capacity distributions that maximize the overall network robustness. 

Our analytical results are supported by extensive numerical simulations with different families of initial load and capacity distributions.  
We show that as the attack size $p$ increases, the system can experience various types of phase transitions, such as continuous (first-order) transitions or a combination of a discontinuous (second-order) transition followed by a continuous transition.  
Ultimately, the system undergoes a complete breakdown through a discontinuous (second-order) transition at a critical attack size.  
In other words, as the attack size increases, the mean fraction of surviving nodes in the steady state undergoes a discontinuous (second-order) transition at the critical attack size, beyond which the system collapses entirely.  
This suggests that cascading failure phenomena, such as those observed in single-layer flow networks like power grids — where initial failures in a small set of nodes propagate through the network and cause a complete breakdown — may also occur in a wider range of systems that align with the \textit{multiplex flow network model}.  

We leverage our results on the solution for the final system size to discuss the optimal configuration of system parameters, such as load and capacity distributions, to maximize robustness against random failures. 
Specifically, we show that allocating total {\em free-space} (i.e., a node's capacity minus initial load) to each layer in proportion to its mean effective load and distributing that free-space equally among nodes within the layer maximizes the system’s critical attack size under the constraint of constant total free space. 
For single-layer flow networks, the \textit{equal free space} allocation, which distributes free space uniformly among all nodes, was shown \cite{single_flow_optimizing} to be {\em optimal} and to significantly outperform the commonly-used \textit{uniform tolerance factor} approach, where the free-space of a node is a constant factor of its initial load~\cite{alpha1,flow_2,alpha2,pei_et.al.,artime_abrupt_2020}. 
Our results extend the \textit{equal free space} strategy to multiplex flow networks through a \textit{layer-weighted equal free space} approach, where free space is first divided among layers according to their mean effective load and then distributed equally among nodes within each layer. 
We also show that in addition to optimizing robustness by attaining the maximum possible critical attack size $p^*$, the \textit{layer-weighted equal free space} approach ensures that for any attack size $p < p^*$, the final system takes its maximum value of $1 - p$, meaning that the this allocation optimizes robustness at all attack sizes $p \in (0,1)$.  
Additionally, we show that the optimal strategy is independent of the probability distribution of its initial load, enhancing its broad applicability.

We believe that the proposed multiplex flow network model, where the layers represent distinct types of flow as opposed to the commonly studied versions where loads are shared across layers \cite{multi_flow_1,scala,zhang_two_flow_redistribution,pei_et.al.}, provides a useful framework that captures a broad range of applications from diverse domains.  
We believe that the formal robustness analysis of the model presented in this work will be helpful in improving the design and control of such systems.  
From a design perspective, this analysis can guide better practices to maximize robustness to cascading failures.  
From a control perspective, insights into the order of phase transitions and the critical attack sizes provide valuable system-level inferences for managing these networks.  

The rest of the paper is organized as follows: In Section~\ref{sec:model}, we define the \textit{multiplex flow network model} and describe the different failure conditions.  
In Section~\ref{sec:main results}, we present the results for the recursive equations, their solution, and the optimality condition.  
These results are verified through numerical simulations in Section~\ref{sec:numerical results}.  
Finally, we discuss the implications of our findings and outline potential future research directions in Section~\ref{sec:discussions and conc}.  

\begin{figure}[!b]
    \centering
    \includegraphics[width=0.95\linewidth]{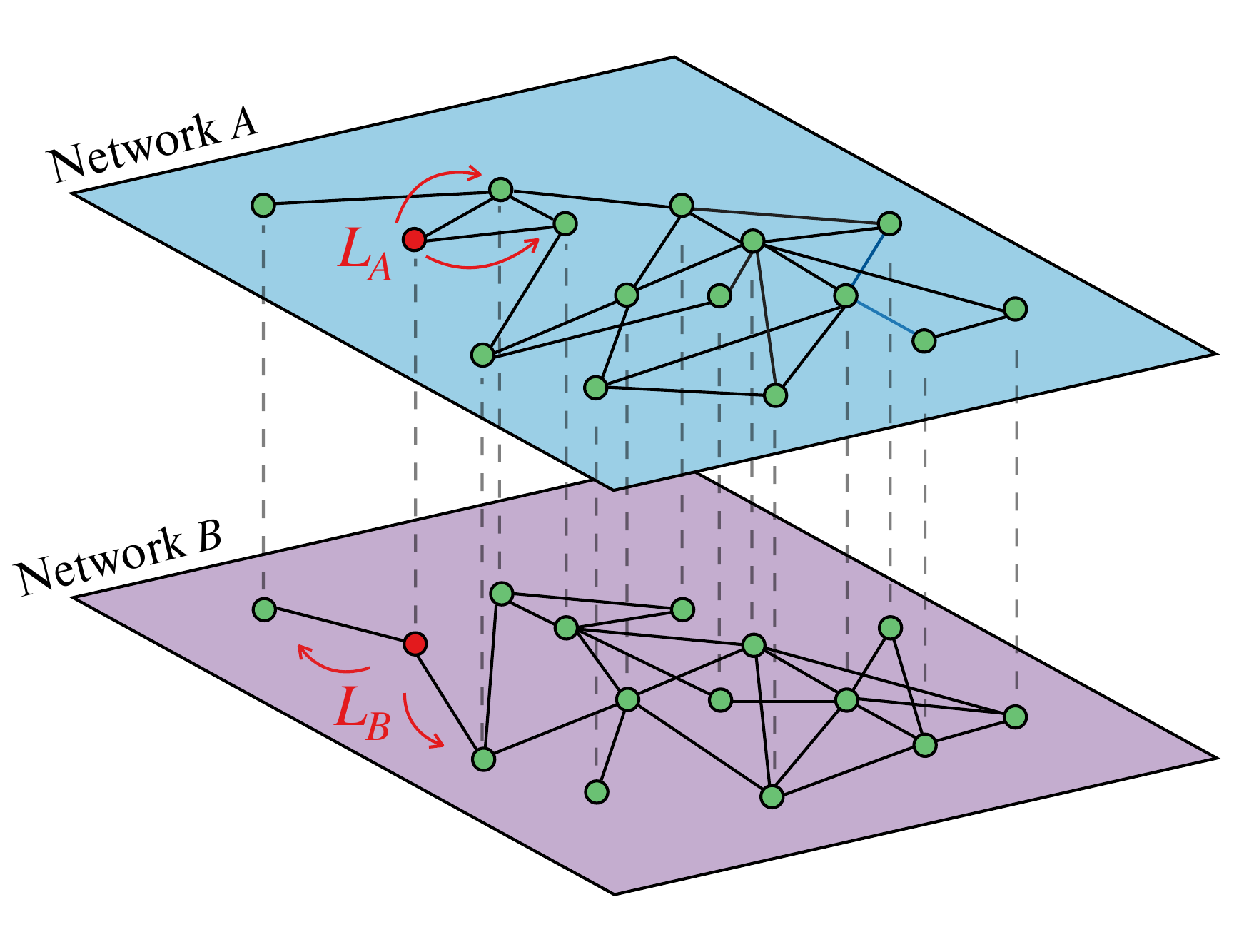}
    \caption{Multiplex Flow Network Model. Each layer is defined on the same set of vertices but with (potentially) different edge sets, and each layer is responsible for carrying/supplying a different flow type. The highlighted node, say $v_x$, carries flow/load of $L_{x,A}$ in network $A$ and $L_{x,B}$ in network $B$; its flow/load vector is then given by $\mathbf{L}_x = [L_{x,A}, L_{x,B}]$. If $v_x$ fails, $L_{x,A}$ amount of flow of type-$A$ will be redistributed to functional nodes in network $A$ and $L_{x,B}$ amount of type-$B$ flow is redistributed to functional nodes in $B$.}
    \label{fig:multiplex_flow_network_fig}
\end{figure}

\section{Multiplex Flow Network Model} 
\label{sec:model}

 The \textit{multiplex-flow network model} is motivated by examples where a set of nodes are responsible from different {\em types} of tasks/functionalities which can be modeled as supporting different types of flow. Consider a set of nodes $\mathcal{N} = \{1,2,...,n\}$ that are responsible for $M\geq2$ distinct types of flows (or, tasks), each of which is transported over (or, supported by) a different graph structure. 
Namely, each flow $\mathcal{M}= \{1,2,...,m\}$ is carried on a graph $G_i$ defined on the vertex set $\mathcal{N}$ and edge set $E_i$. 
This can be represented by a multiplex network with $M$ layers where each layer corresponds to a specific flow type. Figure \ref{fig:multiplex_flow_network_fig} presents an example of the system model with two layers, $\mathcal{M} = \{A,B\}$, representing two distinct modes of flows.
Throughout the paper, we consider the case with  two layers for ease of exposition, but we believe that our results can be extended to cases with $M>2$ in a straightforward manner. 

\vspace{1mm}
\noindent {\bf Overload/Failure Conditions.} In the multiplex flow network model with two layers, the flow/load carried by a node $v_x$ is represented by a two-dimensional vector $\mathbf{L}_x = [L_{x,A},L_{x,B}]$. 
We  define the failure condition (i.e., the overload condition) of a node through a partitioning of the two-dimensional load space into two regions, distinguishing the values of $[L_{x,A},L_{x,B}]$ that the node can sustain from those that would lead the node to be overloaded and fail.
Similarly, the capacities, i.e., the {\em maximum} load that a node can carry, are represented by a two-dimensional vector $\mathbf{C}_x = [C_{x,A},C_{x,B}]$.

\begin{figure}[!t]
    \centering
    \subfigure[Layer-independent Overload]{
            \fbox{\includegraphics[width=0.22\textwidth]{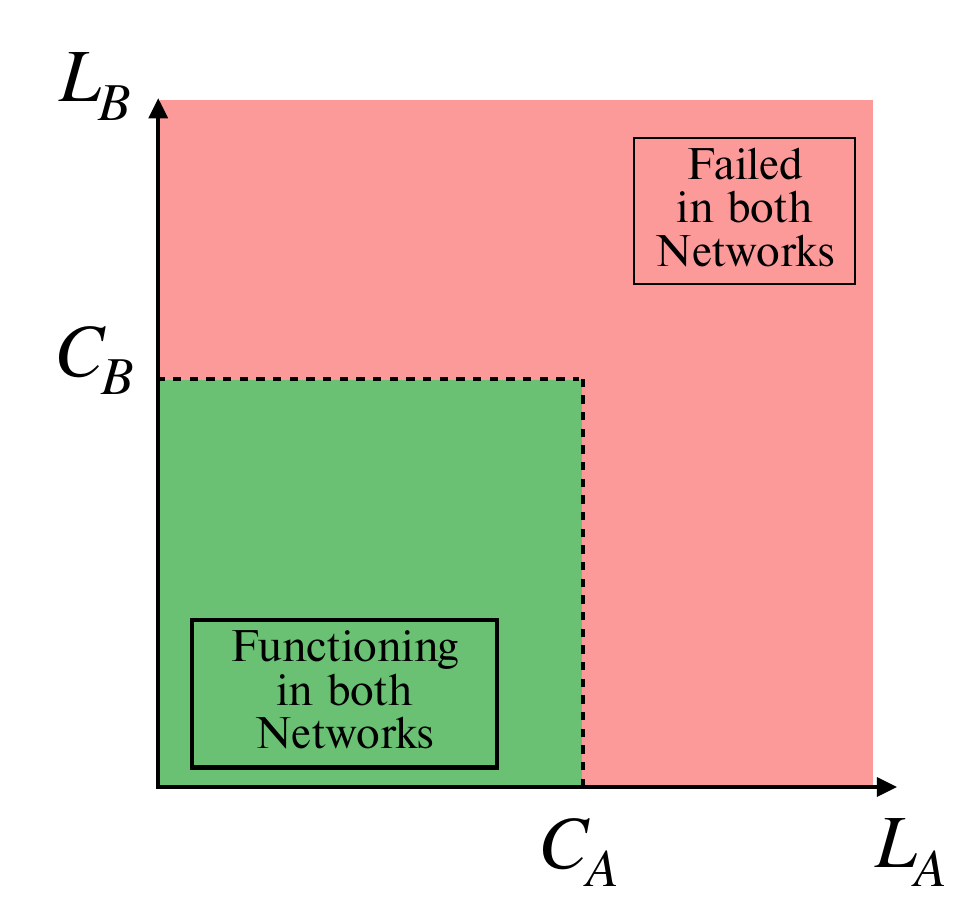}}}
    \subfigure[Layer-influenced Overload]{
            \fbox{\includegraphics[width=0.22\textwidth]{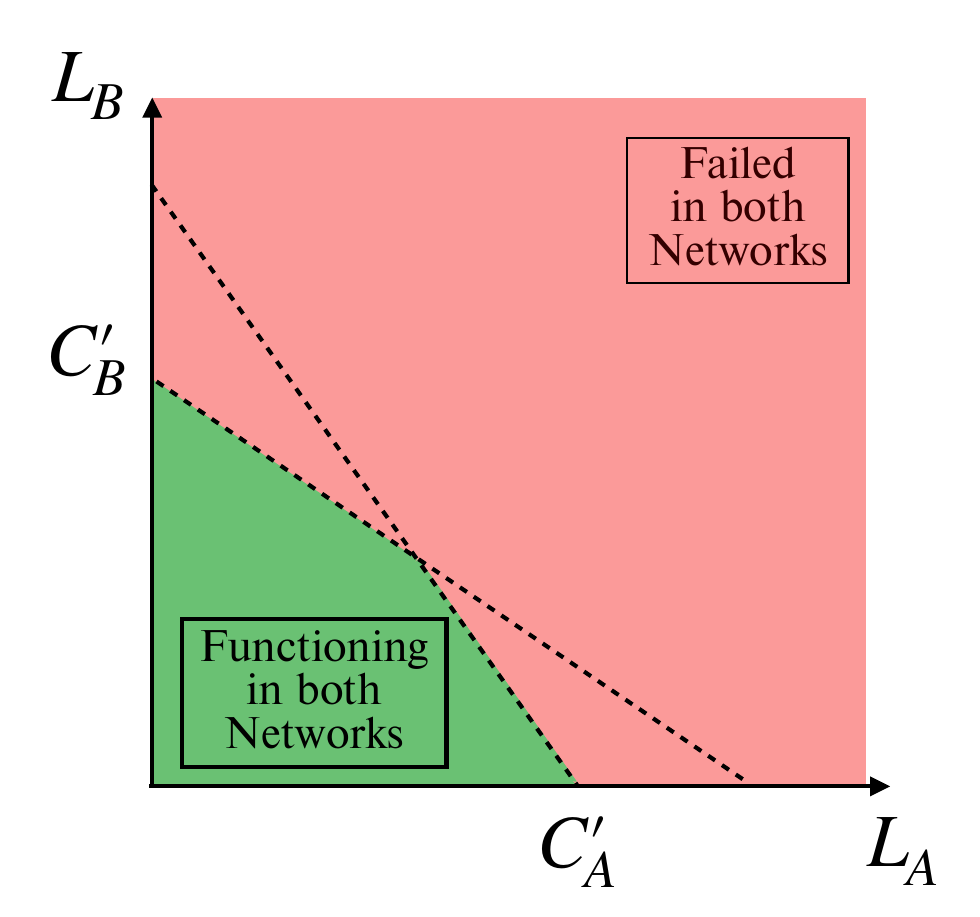}}}   
    \caption{Two different cases for failure/overload conditions defined as a partitioning of the ($L_A, L_B$) plane into regions determining a node’s current state on each flow as functioning or failed. (a) Layer-independent Overload: Independent capacity for each layer and the node fails if capacity is exceeded on any layer. (b) Layer-influenced Overload: Failure condition depends jointly on ($L_A$, $L_B$) indicating that the ability of the node to take on extra load on any given layer is influenced by its load on other layers. Here, we use linear boundaries to define the region where the node is able to function on both layers.}
    \label{fig:failure conditions}
\end{figure}

In terms of specific overload conditions, a simple scenario (referred to as \textit{layer-independent overload} in Figure \ref{fig:failure conditions}) would be when a node keeps functioning as long as its loads in each layer is less than its capacity in the corresponding layer. More formally, the failure condition of node $x$ for \textit{layer-independent overload} can be defined as:
\begin{equation}
    \begin{array}{c}
        L_{x,A} \leq C_{x,A}\text{,} \\
        L_{x,B} \leq C_{x,B}\text{.}
    \end{array}
    \label{eq_main:failure_condition_Case_1}
\end{equation}
Thus, for \textit{layer-independent overload}, the failure condition depends only on the capacity and load within the same layer while overload in one layer causes the failure in function of that node in every layer.

A more complex case worth studying is when the failure condition of a node depends jointly on $\mathbf{L}_x$; e.g., see \textit{layer-influenced overload} in Figure \ref{fig:failure conditions}. 
For this case, the overload condition in a layer is also influenced by the loads in other layer. 
In the case of linear boundaries for the failure regions in the $L_A-L_B$ plane, the surviving condition for \textit{layer-influenced overload} can be formulated as:
\begin{equation}
    \begin{array}{c}
        L_{x,A} + \beta_B L_{x,B} \leq C_{x,A}\text{,} \\
        L_{x,B} + \beta_A L_{x,A} \leq C_{x,B}\text{.}
    \end{array}
    \label{eq_main:failure_condition_Case_2}
\end{equation}

Here $\beta_A$ and $\beta_B$ are the \textit{cross-layer influence factors}, representing the unit impact of the load in one layer on another, for layers $A$ and layer $B$, respectively.
Note that $\beta_A = \beta_B = 0$ corresponds to \textit{layer-independent overload}, where the overload condition in one layer depends only on the load in the same layer. 
Therefore, in Sections \ref{sec:main results}-\ref{sec:discussions and conc}, we present our results for the \textit{layer-influenced overload} case, since setting $\beta_A = \beta_B = 0$ reproduces the results for \textit{layer-independent overload} case.

In addition to the considered cases, many different failure conditions can be defined using the same conceptual model. 
For instance, instead of linear boundaries for \textit{layer-influenced overload}, other studies might consider a different function of $\mathbf{L}_x$ that defines the nature of dependency between layers. 
Alternatively, future work might consider the cases where a node’s state concerning its functionality across layers is decoupled, e.g., a node being able to carry out its functions in layer $A$ but failed in layer $B$, with the overall state still depending jointly on $\mathbf{L}_x$.

\vspace{1mm}
\noindent{\textbf{Flow Redistribution Rule.}} When a node fails, we assume that the flow it was carrying, represented as an $M$-dimensional vector $\mathbf{L}$, is redistributed equally to other functioning nodes in the network.  
Specifically, if a node $v_x$ fails, $L_{x,i}$ units of flow of type-$i$ are redistributed to the rest of the network for each $i \in \mathcal{M}$.  
In this study, we adopt the \textit{global redistribution rule}, where the load of a failed node is redistributed equally among all surviving nodes~\cite{flow_1,scala,fiber_bundle}.  
This approach originates from the \textit{democratic fiber-bundle model} \cite{daniels_jeffreys}, initially developed to study the rupture of fiber bundles under increasing external force.
In the case of \textit{global redistribution rule}, the load of the broken fibers is evenly shared among intact fibers, potentially triggering further ruptures.

The fiber-bundle model (FBM) is particularly useful because it captures the mechanisms of intermittent failures while remaining simple enough to permit analytical solutions in certain cases.
Its general failure-spreading framework makes it applicable to various domains, including traffic networks~\cite{chakrabarti2006fiber}, earthquake dynamics~\cite{earthquake_ex_fiber_bundle}, flow channels~\cite{barre2015cascading}, and electric power systems in high-voltage grids~\cite{flow_3,frasca2021control,fiber_bundle}. 
As opposed to the \textit{global redistribution rule}, future work could also investigate the \textit{local redistribution rule}, where a failing node's load is redistributed to its immediate neighbors only~\cite{pei_et.al.,Chen_Hu_Meng_Yu_2024,hong_suppressing_2016,Wang_Jin_Zhao_2021}.

\vspace{1mm}
\noindent{\textbf{Motivating Applications.}} Although Sections \ref{sec:main results}-\ref{sec:discussions and conc} focus on the analysis of the \textit{layer-influenced overload} condition under the \textit{global load redistribution} rule, the broader conceptual framework of the \textit{multiplex flow network model}, which encompasses various overload conditions and both local and global redistribution rules, applies to a wide range of critical real-world applications. Some examples include:

\begin{itemize}
    \item \textbf{Workload Redistribution in Collaborative Networks:} 
    Companies and government organizations often function as collaborative networks, where individuals work together to complete various tasks.  
    An individual may contribute to multiple tasks, e.g., task-$A$ and task-$B$.  
    If their load vector $[L_A, L_B]$ meets certain failure conditions (e.g., due to  {\em burnout}), they may fail, and their load in terms of both tasks would then be redistributed to other individuals in the networks.  
    For instance, $L_A$ of task-$A$ might be redistributed in network $A$, and similarly for task-$B$ in network $B$.  
    This in turn can trigger additional failures leading to increased workload for remaining nodes and so on, creating a cascade effect. 
    
    \item \textbf{Distributed Computing Infrastructures:}
    Distributed computing infrastructures, such as cloud systems, consist of servers or server clusters with {\em multiple} roles, including web hosting, data storage, and computation.  
    The failure of a cluster might depend on the cumulative burden across various tasks, represented by the load vector $\mathbf{L}$.  
    Upon failure of a cluster, the demand  shifts to other clusters performing similar tasks, potentially leading to resource contention and degraded performance.  
    This may propagate throughout the system, triggering cascading failures.  
    
    \item \textbf{Supply-Chain Networks:}
    Supply chains consist of production and distribution facilities, often responsible for multiple products, e.g., product-$A$ and product-$B$.  
    These systems can be modeled as multiplex networks, where each layer corresponds to a product type, and the topology in each layer reflects facilities covering shared supply and demand.  
    Since products often share resources like raw materials and labor, the production of product-$A$ is influenced by the load from product-$B$, creating interdependencies.  
    When the joint load across both products exceed a node's capability, i.e., the overload condition described in Figure \ref{fig:failure conditions} is satisfied, its production efficiency may decline, leading to unmet demand being redirected to nearby facilities, potentially causing backlogs and lost demands.  
    
    \item \textbf{Material Degradation:}
    The fiber-bundle model \cite{daniels_jeffreys}, commonly used to study material failure under external physical force, can be extended to account for multiple stressors, such as mechanical, chemical, or thermal stress, each represented in a different layer. 
    The load vector $\mathbf{L}$ captures the stress or strain experienced by a material component across these dimensions. 
    Failure conditions in this setting may depend jointly on the loads across all layers, as different types of stress can exhibit intricate dependencies.
    Upon failure, the redistributed stress within each layer may trigger further component failures, potentially leading to cascading effects.
    
    \item \textbf{Financial Networks:} 
    Financial systems are often conceptualized as networks of institutions, such as banks, investment firms, and hedge funds, connected through financial dependencies \cite{fin_netw_1,fin_netw_2}.  
    These institutions engage in various financial instruments, including stocks, bonds, derivatives, and loans, with each instrument forming a distinct network topology.  
    The load vector $\mathbf{L}$ reflects the performance of a node's holdings in different instruments.  
    An institution's survival depends on its overall performance across all instruments, and exceeding a failure threshold (e.g., due to losses) may trigger bankruptcy, adversely affecting the interconnected institutions and causing cascading failures.  
\end{itemize} 

\vspace{1mm}
\noindent{\bf Attack Model and Robustness Metrics.} 
In our analysis, we focus on understanding the cascading failure dynamics initiated by the random removal of a fraction $p$ of nodes in the multiplex network.  
In other words, we consider \textit{random attacks}, where the initially failed set of nodes is selected randomly.  
The initial attack causes a fraction $p \in (0,1)$ of nodes to fail and redistribution of their loads to other functioning nodes.  
Depending on the failure condition, this redistribution can trigger additional overloads and initiate a cascade of failures.  
Instead of \textit{random attacks}, future studies could explore \textit{targeted attacks}, where the failed nodes are selected based on a decision rule, such as prioritizing nodes with the highest loads.  

The main robustness metric that we utilize is the final system size (denoted as $n_\infty(p)$), i.e., the mean fraction of surviving nodes in the steady state as a function of the initial attack size $p$, in the asymptotic limit of large network size $N$.
Let us denote the set of nodes that survived the initial attack $p$ in the steady state as $\mathcal{N}_{\text{surviving}}(p)$.
Then, we define $n_\infty(p)$ as:  
\begin{equation*}  
    n_{\infty}(p) := \lim_{N\to\infty} \frac{\mathbb{E} \left[|\mathcal{N}_{\text{surviving}}(p)|\right]}{N}.  
\end{equation*}  
Throughout the paper, we analyze the final system size $n_{\infty}(p)$ under any attack size $0 < p < 1$.  
A key objective is to determine the \textit{critical attack size}, denoted $p^*$, beyond which the system will fail completely, resulting in $n_\infty(p) = 0$.  
Thus, we formally define $p^*$ as:  
\begin{equation*}  
    p^* := \sup\{p : n_\infty(p) > 0\}.  
\end{equation*}

\section{Main Results}
\label{sec:main results}

In what follows, we present  our main results for \textit{layer-influenced overload} case with linear boundaries, i.e., the overload condition demonstrated in Figure \ref{fig:failure conditions}b.
We consider a two layer network with labels $A$ and $B$. 
Each node is assigned an initial load $[L_{x,A}, L_{x,B}]$ that is identically and independently drawn from given distributions.  
We define the capacities as:
\begin{align*}
      C_{x,A}&=L_{x,A}+\beta_B L_{x,B} + S_{x,A} \\
      C_{x,B}&=L_{x,B}+\beta_A L_{x,A} + S_{x,B}
\end{align*}
where $S_{x,A}$ (respectively,  $S_{x,B}$) denotes the {\em free space} allocated to $v_x$ for flow-$A$ (respectively, flow-$B$).    
Put differently, $S_{x,A}$ indicates the additional flow of type-$A$ that node $v_x$ can handle without failing, and similarly for $S_{x,B}$.
Throughout, we assume that initial load and free space values $(L_{x,A}, S_{x,A}, L_{x,B}, S_{x,B})$ for each node are independently and identically distributed with the joint cumulative distribution function $P_{L_{A}S_{A}L_{B}S_{B}}(y_{L_A},y_{S_A},y_{L_B},y_{S_B})=\mathbb{P}[L_A\leq y_{L_A},S_A\leq y_{S_A},L_B\leq y_{L_B},S_B\leq y_{S_B}]$. 
The corresponding probability density function is denoted by $p_{L_{A}S_{A}L_{B}S_{B}}$. 
To avoid trivial cases, we assume that the joint distributions have positive support, i.e.,  we have $L_{x,A}>0$, $S_{x,A}>0$, $L_{x,B}>0$, $S_{x,B}>0$ for all nodes $v_x$.

In the next three subsections, we present our main results including the recursive equations to calculate the final surviving fraction, the fixed-point solution of the recursion, and the optimal load-free space distribution to maximize robustness for \textit{layer-influenced overload}.

\subsection{Recursive equations for final system size,  $n_\infty(p)$}
Our method for calculating the final system size is based on the recursive calculation of the mean fraction of surviving nodes and the excess load per surviving node induced by the load redistribution.
Each round of flow redistribution corresponds to an iteration $t = 1, 2, 3, \ldots$ during which we first calculate the mean fraction of surviving nodes and then determine the resulting excess load per surviving node.
The cascade begins at $t = 0$, with the initial attack causing a fraction $p$ of the nodes to fail and redistribute their loads.  
The resulting excess load per surviving node in each layer are then calculated and used to compute the fraction of surviving nodes in the next iteration.  
Due to the failures, the set of surviving nodes at iteration $t$, denoted as $\mathcal{N}_t$, is a subset of the surviving nodes from the previous iteration, i.e., $\mathcal{N}_0 \supseteq \mathcal{N}_1 \supseteq \mathcal{N}_2 \supseteq \cdots$.  
We denote the fraction of surviving nodes at iteration $t$ as $n_t$, i.e.,  
\begin{equation*}
    n_t = \frac{|\mathcal{N}_t|}{N} ,\ t=0,1,2, \ldots
\end{equation*}
Moreover, we define the excess load per surviving node at (the end of) iteration $t$ as $Q_{t,A}$ and $Q_{t,B}$ for layers $A$ and $B$, respectively.  
Thus, $Q_{t,A}$ represents the excess load of type $A$ per surviving node, resulting from the redistribution of the loads of failed nodes up to iteration $t$.  
The total excess load per surviving node in each layer is calculated separately for each type of load, $A$ and $B$.  

For the \textit{layer-influenced overload} case with linear boundaries, the failure condition in each layer depends on the loads from the other layer.  
Therefore, the recursive equations must account for the increase in excess loads due to load redistribution in other layers.  
To simplify notation, we define the \textit{effective excess load} in each layer as:  
\begin{align*}  
    Q'_{t,A} = Q_{t,A} + \beta_B Q_{t,B}, \\  
    Q'_{t,B} = Q_{t,B} + \beta_A Q_{t,A}.  
\end{align*}  
Here, $\beta_A$ and $\beta_B$ are the \textit{cross-layer influence factors}, denoting the weights of the load enforced from one layer to the other for layers $A$ and $B$, respectively, with $0 \leq \beta_A, \beta_B$.  
Thus, the \textit{effective excess load} captures the impact of load redistribution on the failure condition in other layers.  
The following result presents how these quantities can be recursively calculated.  

\begin{thm}
\label{thm:recursive equations}
 {\sl   Consider the \textit{multiplex flow network model} for \textit{layer-influenced overload} condition with {\em linear} boundaries under the \textit{global redistribution rule} as described in Section \ref{sec:model}. 
    Assume that the initial load-free space values, $L_{x,A}, S_{x,A}, L_{x,B}, S_{x,B}$, are drawn independently for each node from the joint probability distribution function $p_{L_{A}S_{A}L_{B}S_{B}}$.  
    Let the \textit{cross-layer influence factors} be denoted by $\beta_A$ and $\beta_B$.  
    With $n_0 = 1 - p$, $Q_{0,A} = \frac{p\mathbb{E}[L_A]}{(1-p)}$, and $Q_{0,B} = \frac{p\mathbb{E}[L_B]}{(1-p)}$, the fraction of surviving nodes at each iteration, initiated by a random attack removing a  $p$-fraction of the nodes, can be calculated as follows for each iteration $t = 1, 2, \ldots$  
\begin{align}      
    n_{t} &= n_0\cdot\mathbb{P}[S_A>Q'_{t-1,A},S_B>Q'_{t-1,B}]
    \label{eq_main:final_size_v2} \\
    Q_{t,A} &= \frac{\mathbb{E}[L_A] - (1-p)\mathbb{E}\left[L_A . \mathbbm{1}{\left[\begin{array}{c} S_A > Q'_{t-1,A}\\ S_B > Q'_{t-1,B} \end{array}\right]}\right]}{n_0\cdot\mathbb{P}[S_A>Q'_{t-1,A},S_B>Q'_{t-1,B}]}
    \label{eq_main:excess_load_A_v2} \\
    Q_{t,B} &= \frac{\mathbb{E}[L_B] - (1-p)\mathbb{E}\left[L_B . \mathbbm{1}{\left[\begin{array}{c} S_A > Q'_{t-1,A}\\ S_B > Q'_{t-1,B} \end{array}\right]}\right]}{n_0\cdot\mathbb{P}[S_A>Q'_{t-1,A},S_B>Q'_{t-1,B}]}
    \label{eq_main:excess_load_B_v2}
\end{align}
}
\end{thm}

\begin{table}[htbp]
    \centering
    \begin{tabularx}{\linewidth}{|c|X|}
    \hline
       $\mathcal{N}_t$  & The set of surviving nodes at iteration $t$   \\
    \hline
        $n_t$ &  The fractional size of surviving nodes at iteration $t$ \\
    \hline
        $Q_{t,A}$ &  The excess load of type-$A$ per surviving node at (the end of) iteration $t$ \\
    \hline
        $Q_{t,B}$ &  The excess load of type-$B$ per surviving node at (the end of) iteration $t$ \\
    \hline
        $Q'_{t,A}$ &  The \textit{effective excess load} of type-$A$ per surviving node at (the end of) iteration $t$, i.e., $Q'_{t,A}= Q_{t,A} + \beta_BQ_{t,B}$ \\
    \hline
        $Q'_{t,B}$ &  The \textit{effective excess load} of type-$B$ per surviving node at (the end of) iteration $t$, i.e., $Q'_{t,B}= Q_{t,B} + \beta_AQ_{t,A}$ \\
    \hline
        $\beta_A$ &  The \textit{cross-layer influence factor} representing the unit impact of the load in layer $A$ to layer $B$  \\
    \hline
        $\beta_B$ &  The \textit{cross-layer influence factor} representing the unit impact of the load in layer $B$ to layer $A$  \\
    \hline
    \end{tabularx}
    \caption{Notation used for analysis}
    \label{tab:definitions}
\end{table}

Table \ref{tab:definitions} presents the key notation used in the recursive equations, i.e., (\ref{eq_main:final_size_v2})-(\ref{eq_main:excess_load_B_v2})\footnote{Probabilistic statements are defined using the probability measure $\mathbb{P}$, and the associated expectation operator is represented by $\mathbb{E}$. The indicator function for an event $E$ is written as $\mathbbm{1}[E]$.}.  
In any iteration, if the probability in (3) equals $0$, then $n_t = 0$, and we also set $Q_{t,A} = Q_{t,B} = \infty$, indicating the total collapse of the system.  
Below, we give a brief outline of the proof of Theorem \ref{thm:recursive equations}, while detailed derivations are presented in 
 Appendix \ref{sec:apx_final size}.

The recursive equations are obtained using the \textit{mean-field analysis} of cascading dynamics for the \textit{multiplex flow-network model} under the \textit{layer-influenced overload} case.  
This implies that the calculations provided for $n_t$ reflect the mean fraction of surviving nodes in the asymptotic limit of a large network size $N$.  
To derive the equations, we first calculate the mean fraction of surviving nodes and the resulting excess loads after the initial attack.  
An initial attack of size $p$ at iteration $t = 0$ results in a system size of $n_0 = 1 - p$.  
To compute the mean excess loads per surviving node in each layer, we consider the loads redistributed from the failures and divide them by the relative size of the surviving nodes.  
Since the initial attack is random, the mean excess loads per surviving node, $Q_{0,A}$ and $Q_{0,B}$, can be easily calculated as $\frac{E[L_A]p}{1 - p}$ and $\frac{E[L_B]p}{1 - p}$, respectively.  
We begin with iteration $t = 1$ and recursively calculate the mean fraction of surviving nodes and the excess loads per surviving node for $t = 2, 3, \ldots$, to obtain expressions for $n_t$, $Q_{t,A}$, and $Q_{t,B}$ in terms of their values at iteration $t-1$.

The recursive relationship between $n_{t+1}$ and $n_{t}$ is established based on the fact that, at each iteration, only the nodes with capacities larger than their effective loads will survive.  
In other words, the nodes in $\mathcal{N}_t$ with free space smaller than $Q'_{t,A}$ and $Q'_{t,B}$ will fail at iteration $t+1$.  
Thus, at every iteration, a fraction of the surviving nodes will progress to the next iteration, depending on the probability of having free space larger than the current effective excess load per surviving node.  
Note that, at iteration $t$, nodes with less free space than the effective excess loads from the previous iteration, $Q'_{t-1,A}$ and $Q'_{t-1,B}$, have already failed.  
As a result, the probability of failure at this iteration must be conditioned on the nodes that have survived up to that point.  
This leads to:  
\begin{equation}  
    n_{t+1} = n_t \cdot \mathbb{P}\left(\begin{array}{c|c} S_A > Q'_{t,A} & S_A > Q'_{t-1,A} \\ S_B > Q'_{t,B} & S_B > Q'_{t-1,B} \end{array}\right).  
    \label{eq_main:final_size_v1}  
\end{equation}  

The conditional probability in (\ref{eq_main:final_size_v1}) corresponds to the survival probability of a node that have already survived until $t - 1$ and will survive in the next iteration.  
Moreover, as we progress through the iterations, the effective excess loads must either remain constant, in the case of no additional failures, or increase due to the load redistribution from newly failed nodes.  
This implies that $Q'_{t,A} \geq Q'_{t-1,A}$ and $Q'_{t,B} \geq Q'_{t-1,B}$, allowing us to express the conditional probability in (\ref{eq_main:final_size_v1}) as the ratio: $\frac{\mathbb{P}[S_A>Q'_{t,A},S_B>Q'_{t,B}]}{\mathbb{P}[S_A>Q'_{t-1,A},S_B>Q'_{t-1,B}]}$.
We then apply repeated substitutions for $n_t$, $n_{t-1}$, $\ldots$, $n_0$ to derive the final form provided in (\ref{eq_main:final_size_v2}) (see Appendix \ref{sec:apx_final size} for details).  

Equations for calculating the excess loads per surviving node, $Q_{t,A}$ and $Q_{t,B}$, are derived by computing the total excess loads (of type-$A$ and type-$B$) to be redistributed among all remaining nodes up to iteration $t$ and dividing it by the relative size of the surviving nodes.  
To this end, the total excess load redistributed until $t$ can be decomposed into two components: the failures resulting from the initial attack, and the failures caused by cascading effects initiated by the initial attack.  
This total excess load is then divided by the relative size of the surviving nodes to obtain the excess load per surviving node.  
This relationship can be seen more clearly when we rearrange the expression in (\ref{eq_main:excess_load_A_v2}) by substituting $\mathbb{E}[L_A]$ with $p\mathbb{E}[L_A] + (1 - p)\mathbb{E}[L_A]$ and replacing the denominator with $n_{t-1}$ using (\ref{eq_main:final_size_v2}), which yields:  
\begin{align}
    \nonumber
    &Q_{t,A} = \\ &\frac{p\mathbb{E}[L_A] + (1 - p)\mathbb{E}\left[L_A \left( 1 - \mathbbm{1}{\left[\begin{array}{c} S_A > Q'_{t-1,A} \\ S_B > Q'_{t-1,B} \end{array}\right]}\right)\right]}{n_{t}}  
    \label{eq_main:excess_load_A_v3}  
\end{align}  
Note that the first term in the numerator of (\ref{eq_main:excess_load_A_v3}) represents the excess load caused by the initial attack, normalized by the total number of nodes ($N$), while the second term accounts for the total excess load resulting from cascading failures that occur after the initial attack, also divided by $N$.  
Finally, the expression is divided by the fractional size of the surviving nodes in the denominator   
to convert the total excess load normalized by $N$ into the total excess load \textit{per surviving node}.

In summary, (\ref{eq_main:final_size_v2})-(\ref{eq_main:excess_load_B_v2}) enable calculating the final system size by iteratively computing $n_t$ until the system reaches a {\em steady state}, i.e., to a point where $n_{t+1} = n_t$ and no further failures or redistribution take place.
The final value obtained, i.e, $n_{t+1} = n_t$, corresponds to the final system size.  
In the next subsection, we explore {\em fixed-point} solutions of the recursion $n_{t+1} = n_t$  that allow computing the final system size $n_\infty(p)$ directly without iterating over (\ref{eq_main:final_size_v2})-(\ref{eq_main:excess_load_B_v2}) for $t=1, 2, \ldots$.

\subsection{Solving the recursions for calculating the final system size}
In the previous section, we derived the recursive equations presented in the previous section enable calculating the final system size $n_\infty(p)$ iteratively through the recursions (\ref{eq_main:final_size_v2})-(\ref{eq_main:excess_load_B_v2}).  
Our next result shows how the final system size can be computed {\em without} going through the iterations and instead finding a the fixed-point solutions to $n_{t+1} = n_t$. 

\begin{thm}
\label{thm:solution set}
{\sl
    Let $\mathcal{P}$ denote the set of ``\textit{stable points}'', defined by the condition:
    \begin{equation}
    \mathcal{P}:\{x,y: x\geq g(x,y),y\geq h(x,y)\}
    \label{eq_main:solution set}
    \end{equation}
    where
    \begin{equation*}
    g(x,y) = \frac{\mathbb{E}[L_A] - (1-p)\mathbb{E}\left[L_A . \mathbbm{1}{\left[\begin{array}{c} S_A > x + \beta_B y\\ S_B > y + \beta_A x 
             \end{array}\right]}\right]}{(1-p).\mathbb{P}[S_A>x + \beta_B y,S_B>y + \beta_A x]}
    \end{equation*}
    \begin{equation*}
    h(x,y) = \frac{\mathbb{E}[L_B] - (1-p)\mathbb{E}\left[L_B . \mathbbm{1}{\left[\begin{array}{c} S_A > x+ \beta_B y\\ S_B > y + \beta_A x 
             \end{array}\right]}\right]}{(1-p)\mathbb{P}[S_A>x+ \beta_B y,S_B>y+ \beta_A x]}
    \end{equation*}
    \begin{itemize}
    \item[i)]  $\mathcal{P}$ has an element-wise minimum point, i.e., there exists $(x^*,y^*)\in \mathcal{P}$ such that $x^*\leq x \text{ and } y^* \leq y \text{ for all } (x,y)\in \mathcal{P}.$
    \item[ii)] The mean fraction of surviving nodes at the steady state, i.e., the final system size, 
    can be calculated as:
    \begin{align}
       \nonumber 
    & n_{\infty}(p) 
    \\ & = (1-p) \mathbb{P}[S_A>x^*+\beta_B y^*,S_B>y^*+\beta_A x^*]. \label{eq: final size}
    \end{align}
    \end{itemize}
    }
\end{thm}

A detailed proof of Theorem \ref{thm:solution set} is provided in Appendix~\ref{sec:apx_fixed point solution}. Here, we provide an overview of the intuition behind this result.  
In view of (\ref{eq_main:final_size_v1}), to reach a steady-state, i.e., to have $n_{t+1} = n_t$, we must have $\mathbb{P}(S_A > Q'_{t,A}, S_B > Q'_{t,B} \mid S_A > Q'_{t-1,A}, S_B > Q'_{t-1,B}) = 1$.
In other words, the problem of finding the fixed-point solutions satisfying $n_{t+1} = n_t$ is equivalent to identifying excess load values $(x, y) = (Q_{t,A}, Q_{t,B})$   that  do not lead to further failures. 
Intuitively, the cascading failures will stop and the final system size can be computed at the {\em first} such fixed-point solution that the system reaches.  
In Appendix~\ref{sec:apx_fixed point solution}, we present two key technical results to characterize this solution. 
\textit{Lemma}~\ref{lemma:Claim 1}  shows that when searching for the pair $(Q_{t,A}, Q_{t,B})$ that satisfies the stopping condition $n_{t+1} = n_t$, it is sufficient to consider the set of points satisfying $Q_{t,A} \geq Q_{t+1,A}$ and $Q_{t,B} \geq Q_{t+1,B}$, which are defined as \textit{stable points}.  \textit{Lemma}~\ref{lemma:Claim 2} establishes that, within the set of \textit{stable points} $\mathcal{P}$, there exists an element-wise minimum point $(x^*, y^*)$ such that $x^* \leq x$ and $y^* \leq y$ for all $(x, y) \in \mathcal{P}$. In other words, $(x^*, y^*)$ will be the {\em first} stable point that the system will reach, giving the {\em smallest} values of  excess loads  $(Q_{t,A}, Q_{t,B})$ that will not lead to any further failures in the system. Thus, the final system size can be computed, as given in (\ref{eq: final size}), as the fraction of nodes who are not failed at the initial attack {\em and} who have enough free space to handle the excess loads of $(x^*, y^*)$.

\begin{figure*}[b]
    \centering
    \subfigure[Condition for $A$]{
            \fbox{\includegraphics[width=0.25\textwidth,trim={2cm 1cm 1cm 2.5cm},clip]{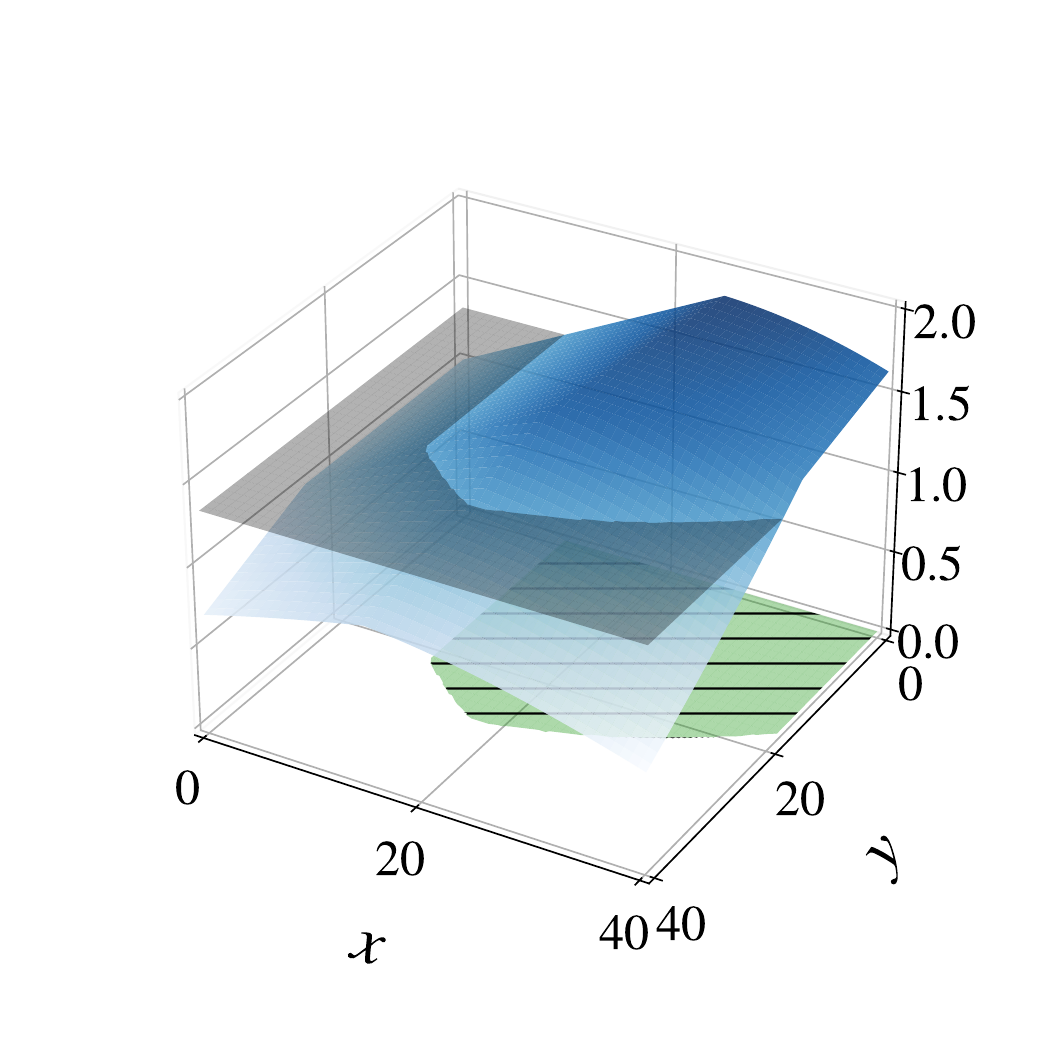}}}
    \subfigure[Condition for $B$]{
            \fbox{\includegraphics[width=0.25\textwidth,trim={2cm 1cm 1cm 2.5cm},clip]{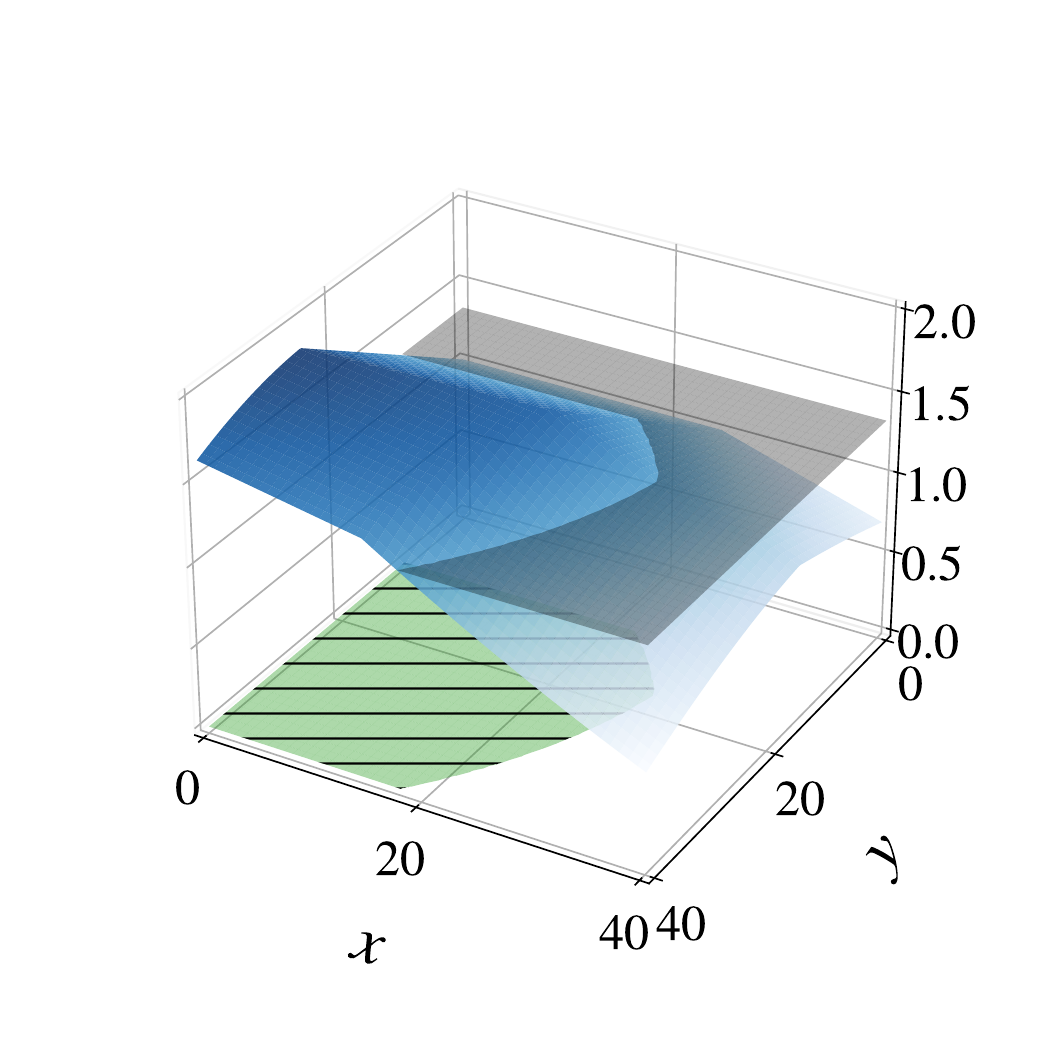}}}
    \subfigure[Stable Points]{
            \fbox{\includegraphics[width=0.25\textwidth,trim={2cm 1cm 1cm 2.5cm},clip]{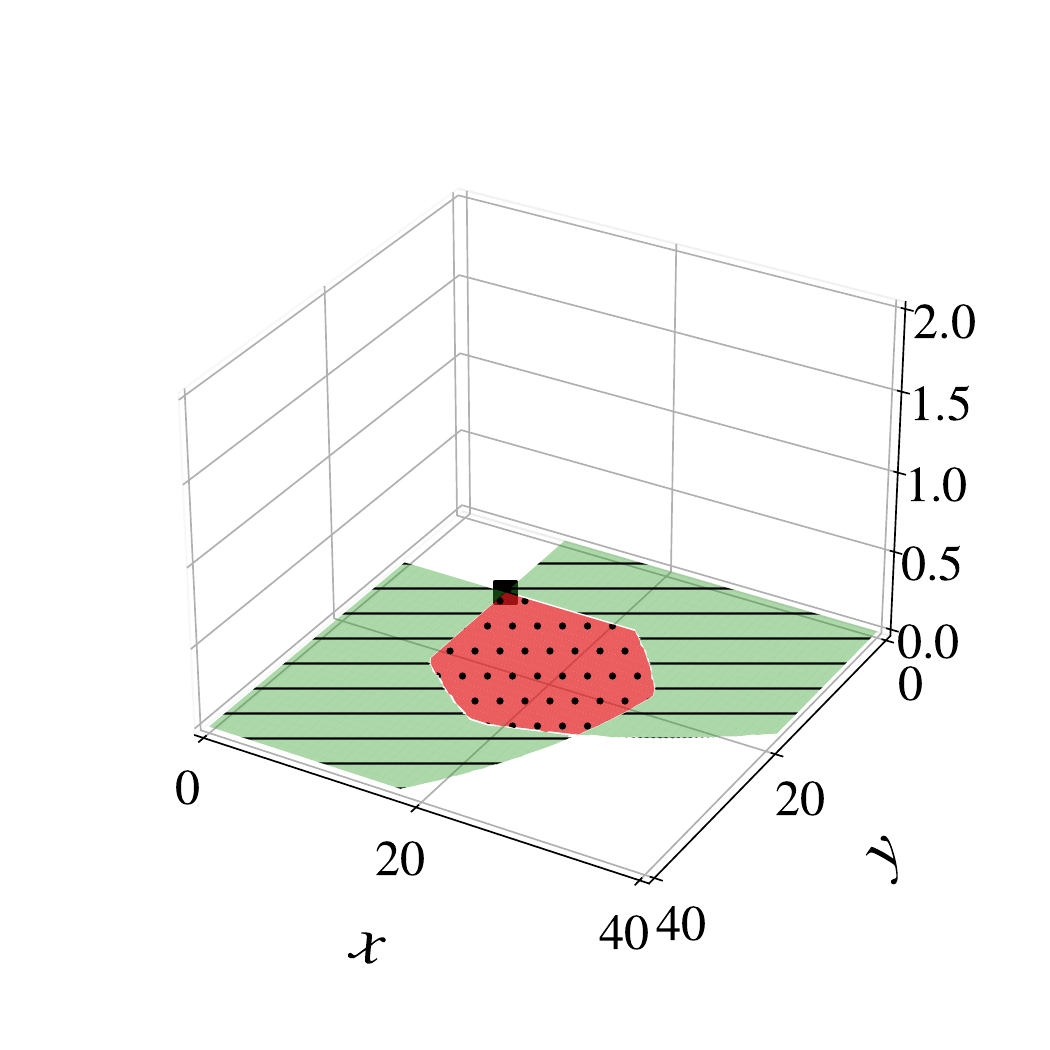}}}\\
    \subfigure{
        \fbox{\includegraphics[width=0.75\textwidth,trim={0.5cm 2cm 1cm 2cm},clip]{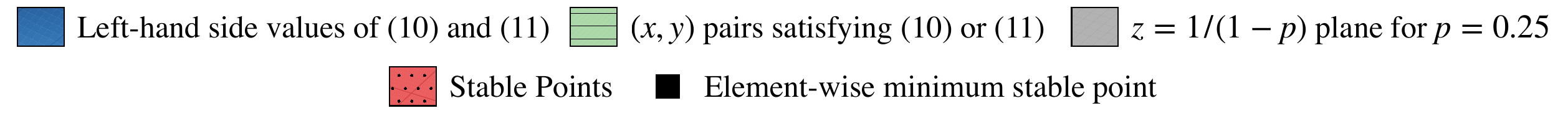}}}
    \caption{The resulting plots illustrating the graphical solution procedure. The initial loads in both layers are distributed uniformly with $L_{\textrm{min}} = 20$ and $\mathbb{E}[L] = 30$, while the free spaces in both layers are also distributed uniformly with $S_{\textrm{min}} = 25$ and $\mathbb{E}[S] = 50$. Additionally, $\beta_A = \beta_B = 0.25$. Subfigures (a) and (b) depict the surviving conditions for Networks $A$ and $B$, where the blue surface represents the left-hand side values of the inequalities. The gray plane (parallel to the \textit{x-y} plane) corresponds to $z =1/(1-p)$ when $p = 0.25$, and the $(x,y)$ values above this plane are projected onto the \textit{x-y} plane in green (horizontally striped area). Figure (c) shows the intersection of points that satisfy both inequalities, referred to as \textit{stable points}, highlighted in red (dotted area). The upper-left corner of the stable points (black square), indicating the element-wise minimum, represents the resulting excess load for the given attack size at steady state, corresponding to a surviving fraction of $0.75$. As $p$ increases, the gray plane shifts upwards, shrinking the set of stable points. Eventually there will be a critical attack size $p^*$ beyond which we observe complete failure of the system indicated by the lack of feasible solutions for the stable points.  }
    \label{fig:graphical Solution}
\end{figure*}

\subsubsection*{\textbf{Graphical Interpretation}}

The solution set presented in \textit{Theorem}~\ref{thm:solution set} provides a way to interpret how a system with given initial load and free space distributions, along with the \textit{cross-layer influence factors} ($\beta_A,\beta_B$), will behave for a specified initial attack size. 
Considering the inequalities in (\ref{eq_main:solution set}), if we explicitly write the functions $g(x,y),h(x,y)$, we can rewrite the conditions $x\geq g(x,y),y\geq h(x,y)$ as the following two inequalities:
\begin{align}
\nonumber
\frac{\mathbb{P}\left[ \begin{array}{c} S_A > x + \beta_B y\\ S_B > y + \beta_A x 
         \end{array} \right] \left( x + \mathbb{E} \left[L_A\ \begin{array}{|c} S_A > x + \beta_B y\\ S_B > y + \beta_A x 
         \end{array} \right] \right)}{\mathbb{E}[L_A]} & \\
 & \hspace{-7.5cm} \geq  \frac{1}{1-p}
         \label{layer_ineq_A}\\
\nonumber
\frac{\mathbb{P}\left[ \begin{array}{c} S_A > x + \beta_B y\\ S_B > y + \beta_A x 
         \end{array} \right] \left( y + \mathbb{E} \left[L_B \ \begin{array}{|c} S_A > x + \beta_B y\\ S_B > y + \beta_A x 
         \end{array} \right] \right)}{\mathbb{E}[L_B]} & \\
    & \hspace{-7.5cm}       \geq \frac{1}{1-p}&& 
         \label{layer_ineq_B}
\end{align}
(\ref{layer_ineq_A}) and (\ref{layer_ineq_B}) corresponds to the survival condition in layers $A$ and $B$, respectively. 
The intersection of $(x, y)$ points that satisfy both inequalities determines the set of \textit{stable points}. 
Among these, the element-wise minimum stable point represents the final excess loads in each layer, which can then be used to calculate the final fraction of surviving nodes.

Note that (\ref{layer_ineq_A}) and (\ref{layer_ineq_B}) do not have a term depending on $p$ on the left-hand side. 
This implies that, given the initial load/free space distributions and the $\beta$ values, we can compute the left-hand side values for a range of $x$ and $y$ and represent them as two surfaces in a 3D plot. 
Here, $x$ and $y$ correspond to the excess load values in networks $A$ and $B$, respectively, and should therefore be non-negative values.

Figure \ref{fig:graphical Solution} illustrates the resulting plots for this procedure, using uniformly distributed initial load and free space values with $L_{A,\text{min}} = L_{B,\text{min}} = 20$, $\mathbb{E}[L_A] = \mathbb{E}[L_B] = 30$, $S_{A,\text{min}} = S_{B,\text{min}} = 25$, $\mathbb{E}[S_A] = \mathbb{E}[S_B] = 50$, and $\beta_A = \beta_B = 0.25$. 
In Figure \ref{fig:graphical Solution}(a) and (b), the left-hand side values for $(x, y)$ are shown in blue. The gray plane represents the $z = 1/(1-p)$ plane, with $p = 0.25$. The $(x, y)$ pairs satisfying the inequality are projected onto the \(x\)-\(y\) plane in green (horizontally striped area) for Figure \ref{fig:graphical Solution}(a) and (b). 
Finally, Figure \ref{fig:graphical Solution}(c) shows the set of stable points in red (dotted area), which satisfy both conditions. 
The element-wise minimum of these points (denoted as black square) should be identified and used to calculate the final fraction of surviving nodes.

Considering the gray plane indicating $z = 1/(1-p)$, we observe that as $p$ increases, the plane will be shifting upwards, shrinking the set of stable points and eventually leaving no stable points. 
Thus, the last value of $p$ that still results in a non-empty set of stable points will correspond to the \textit{critical attack size}, $p^*$ which is defined as the maximum attack size, beyond which any attack will cause complete failure. 

The graphical analysis thus enables us to obtain the critical attack size for a given system or calculate the final fraction of surviving nodes under any attack size of $p$. 
Although less intuitive, further analysis on the order of phase transitions can also be performed using additional representations in the same plots.
However, we omit such analysis here, as the interpretability of 3D plots becomes increasingly limited with the addition of more elements.

\subsection{Optimizing Robustness}

From the perspective of \textit{designing} a system, it is crucial to determine the optimal load and free space distributions that maximize robustness under given constraints. 
In many real-world scenarios, resources are inherently limited, making it meaningful to analyze systems where the capacity —and by extension, free space— is constrained. 
Accordingly, we consider a configuration in which the total expected free space across the layers of the multiplex network remains constant. 
In other words, we consider the case $\mathbb{E}[S_A] + \mathbb{E}[S_B] = S_{\text{total}}$, where $S_{\text{total}}$ is a constant.
The next result shows how the system robustness can be optimized, both in terms of the critical attack size $p^*$ and the final system size, $n_\infty(p)$ for all $p \in (0,1)$,  under this constraint by appropriate selection of the free space distributions. 
\begin{thm}
    \label{thm: Optimality}
  {\sl   Consider the \textit{multiplex flow network model} described in \textit{Theorem}~\ref{thm:recursive equations}.  
    Let $\mathbb{E}[S_A] + \mathbb{E}[S_B] = S_{\text{total}}$ where $S_\text{total}$ is a constant.
    \begin{itemize}
    \item[i)] Under these constraints, the critical attack size $p^*$ is upper-bounded as  $p^* \leq p^*_\text{opt}$, where
    \begin{align}
    \label{eq_main:critical_attack_size_opt}
    & p_{opt}^*  
    \\ 
    & =  \frac{\mathbb{E}[S_A] +\mathbb{E}[S_B]}{\mathbb{E}[S_A] + \mathbb{E}[S_B]+(1+\beta_A)\mathbb{E}[L_A] + (1+\beta_B) \mathbb{E}[L_B]} 
   \nonumber
    \end{align}
    \item[ii)]
    The optimal critical attack size,   $p^*_\text{opt}$,
 can be achieved by choosing the free space distributions as
    \begin{align*}
        p_{S_A} &= \delta\left(x-S_{\text{total}}\frac{\mathbb{E}[L_A]+\beta_B \mathbb{E}[L_B]}{(1+\beta_A)\mathbb{E}[L_A]+(1+\beta_B)\mathbb{E}[L_B]}\right)\\
        p_{S_B} &= \delta\left(x-S_{\text{total}}\frac{\mathbb{E}[L_B]+\beta_A \mathbb{E}[L_A]}{(1+\beta_A)\mathbb{E}[L_A]+(1+\beta_B)\mathbb{E}[L_B]}\right)
    \end{align*}   
    irrespective of how initial loads are distributed. 
   \item[iii)] The free space distributions given above maximizes the final system size, $n_{\infty}(p)$ for all $p$ in $(0,1)$. Namely, with $n_{\infty,\text{Dirac}}$ denoting the final system size under this configuration, we have:
    \begin{equation*}
        n_{\infty,\text{Dirac}}= 
        \begin{cases}
            1-p,&  p\leq p^*_\text{opt}\\
            0,              & p>p^*_\text{opt}
        \end{cases}
    \end{equation*}
    \end{itemize}
}
\end{thm}

A proof of  Theorem \ref{thm: Optimality} is provided in Appendix~\ref{sec:apx_optimality}.  
In the proof, we first consider the upper bounds for the critical attack sizes of individual layers $A$ and $B$ by ignoring the dependency between layers, for general initial distributions for $L$ and $S$.  
We then utilize the fact that the critical attack size of the two-layer multiplex network with $\mathbb{E}[S_A] + \mathbb{E}[S_B] = S_{\text{total}}$ is bounded by the critical attack sizes of the individual networks.  
This is because, in the \textit{multiplex flow network model}, the complete failure of one layer results in the complete failure of the system.  
Using this approach, we derive the first part of our result, given by (\ref{eq_main:critical_attack_size_opt}), which defines the upper bound of the critical attack size for any system with $\mathbb{E}[S_A] + \mathbb{E}[S_B] = S_{\text{total}}$.  
Next, using the recursive equations (\ref{eq_main:final_size_v2})-(\ref{eq_main:excess_load_B_v2}), we show that the proposed initial free space distribution will not cause additional failures after an initial attack of size $[p^*_\text{opt}]^-$.  
This implies that, for any attack size $p < p^*_\text{opt}$, the system reaches a final size of $1 - p$.  
This result is significant because the proposed optimal configuration not only achieves the upper bound $p^*_\text{opt}$, but also ensures the maximum possible final system size of $1 - p$ following a random failure of size $p$.  

The proposed initial free space distribution strategy in \textit{Theorem}~\ref{thm: Optimality} suggests that, to maximize robustness, the total free space should be distributed across each layer based on their expected loads, adjusted by the \textit{cross-layer influence factors}: $\beta_A,\beta_B$.  
Subsequently, the allocated free space within each layer should be evenly distributed among all nodes.  
For example, if the mean effective load in layer $A$ is twice that in layer $B$, then two-thirds of the total free space would be allocated to layer $A$, and one-third to layer $B$.  
The allocated free space is then equally distributed among all nodes within each layer.  
This strategy is consistent with previous work on single-layer flow networks \cite{single_flow_optimizing}, which suggests that \textit{equal free space} allocation—distributing the total free space uniformly across all nodes—results in maximum robustness.  
In this sense, our theorem generalizes the optimal strategy of \textit{equal free space} allocation from single-layer flow networks to multiplex flow networks, with the initial step of \textit{weighted} allocation of $S_{\text{total}}$ \textit{to each layer}.  
Hence, we refer to this strategy as ``\textit{layer-weighted equal free space}'' throughout the rest of the paper.

Instead of the total free space constraint considered in Theorem~\ref{thm: Optimality}, a related optimization question arises when the expected free spaces in each layer are fixed. Namely, we might be given that \( \mathbb{E}[S_A]=\mu_A \) and \( \mathbb{E}[S_B]=\mu_B \), where \( \mu_A, \mu_B \) are constants.  
In this scenario, since the total expected free space given to each layer is predetermined, the key question is how to allocate this free space among individual nodes. 
In Appendix~\ref{sec:apx_optimality} (Theorem~\ref{thm: Optimality_v2}), we show that \textit{equal free space} allocation—distributing the total  expected free space uniformly among all nodes—also yields the optimal critical attack size and final system size. Namely, the allocation where each node is given a free space of $\mu_A$ in layer-$A$ and a free space of $\mu_B$ in layer-$B$ leads to maximum robustness under the constraints 
that $\mathbb{E}[S_A]=\mu_A $ and $ \mathbb{E}[S_B]=\mu_B$.  

The proof follows the same steps as in Theorem~\ref{thm: Optimality}, with the only difference being that the critical attack size of the system is bounded by the critical attack size of the weakest layer in terms of robustness, given \( \mu_A, \mu_B \).  
Specifically, we first compute the maximum critical attack sizes of individual layers, denoted as \( p_A \) and \( p_B \), assuming no interdependence between layers.  
We then demonstrate that the equal free space configuration for the multiplex case attains the critical attack size of the weakest layer, i.e., \( p^*_{\text{opt}}=\min(p_A,p_B) \), while also maximizing the final system size, given by \( n_{\infty,\text{Dirac}}=1-p \) for \( p \leq p^*_{\text{opt}} \).

\section{Numerical Results}
\label{sec:numerical results}
In this section, we present numerical results validating the analytical results presented earlier concerning  the recursive equations for the final system size, the fixed-point solution of the recursions, and the optimal allocation of free space.  To this end, we conducted numerical simulations with different distributions for initial load and free space values $L_A,L_B,S_A,S_B$. More specifically, we used the following distributions:
\begin{itemize}
    \item Uniform Distribution: $L \sim U(L_{\textrm{min}},L_{\textrm{max}})$ with probability density function given by
\begin{equation*}
        p_{L}(x)= \frac{1}{L_{\textrm{min}} - L_{\textrm{max}}}  \mathbbm{1} [L_{\textrm{min}}\leq x \leq L_{\textrm{max}}].
\end{equation*}

    \item Pareto Distribution: $L \sim \textrm{Pareto}(L_{\textrm{min}},b)$. With $L_{\textrm{min}}>0$ and $b>0$, the  probability density function  is given by
\begin{equation*}
        p_{L}(x)= L_{\textrm{min}}^bbx^{-b-1} \mathbbm{1} [L_{\textrm{min}}\leq x].
\end{equation*}

    \item Weibull Distribution: $L \sim \textrm{Weibull}(L_{\textrm{min}},\lambda,k)$. With $\lambda,k,L_{\textrm{min}}>0$, the probability density  function is given by
    \begin{equation*}
                p_{L}(x)= \frac{k}{\lambda} \left( \frac{x-L_{\textrm{min}}}{\lambda}\right)^{k-1}e^{-\left(\frac{x-L_{\textrm{min}}}{\lambda}\right) ^ k} \mathbbm{1} [L_{\textrm{min}}\leq x].
    \end{equation*}
        
\end{itemize}

These distributions provide a comprehensive characterization of different load conditions commonly encountered in real-world systems.  
The uniform distribution serves as a natural baseline, representing scenarios where all load or free space values are equally likely. 
The heavy-tailed nature of the Pareto distribution, which allows for extreme values, captures scenarios where a few nodes may have significantly large load or free space values.  
The Weibull distribution is capable of representing a broad range of statistical distribution families, such as the Exponential or Rayleigh distributions, and thus has wide applicability in workload analysis, particularly in modern engineering systems, such as computing workloads in distributed systems~\cite{cloud_comp_weib}.  
Together, these distributions span diverse load characteristics, making the evaluation thorough and representative of practical scenarios.

It is also important to note that, although the results presented in this section are obtained when load distributions  are independent across different layers (i.e., $L_A$ and $L_B$ are independent), we 
also  validated that our findings do not depend on the assumption of independence among these random variables.
To this end, we
also conducted experiments using a Multivariate Normal Distribution for $L_A$ and $L_B$ with non-zero covariance.  
This approach does not assume independence among $L_A$, $L_B$, $S_A$, and $S_B$, and it still produces results consistent with the  simulations with independent distributions.

\subsection{Numerical Validation of Recursive Equations}
\begin{figure}[!b]
    \centering
    \includegraphics[width = 0.98\linewidth]{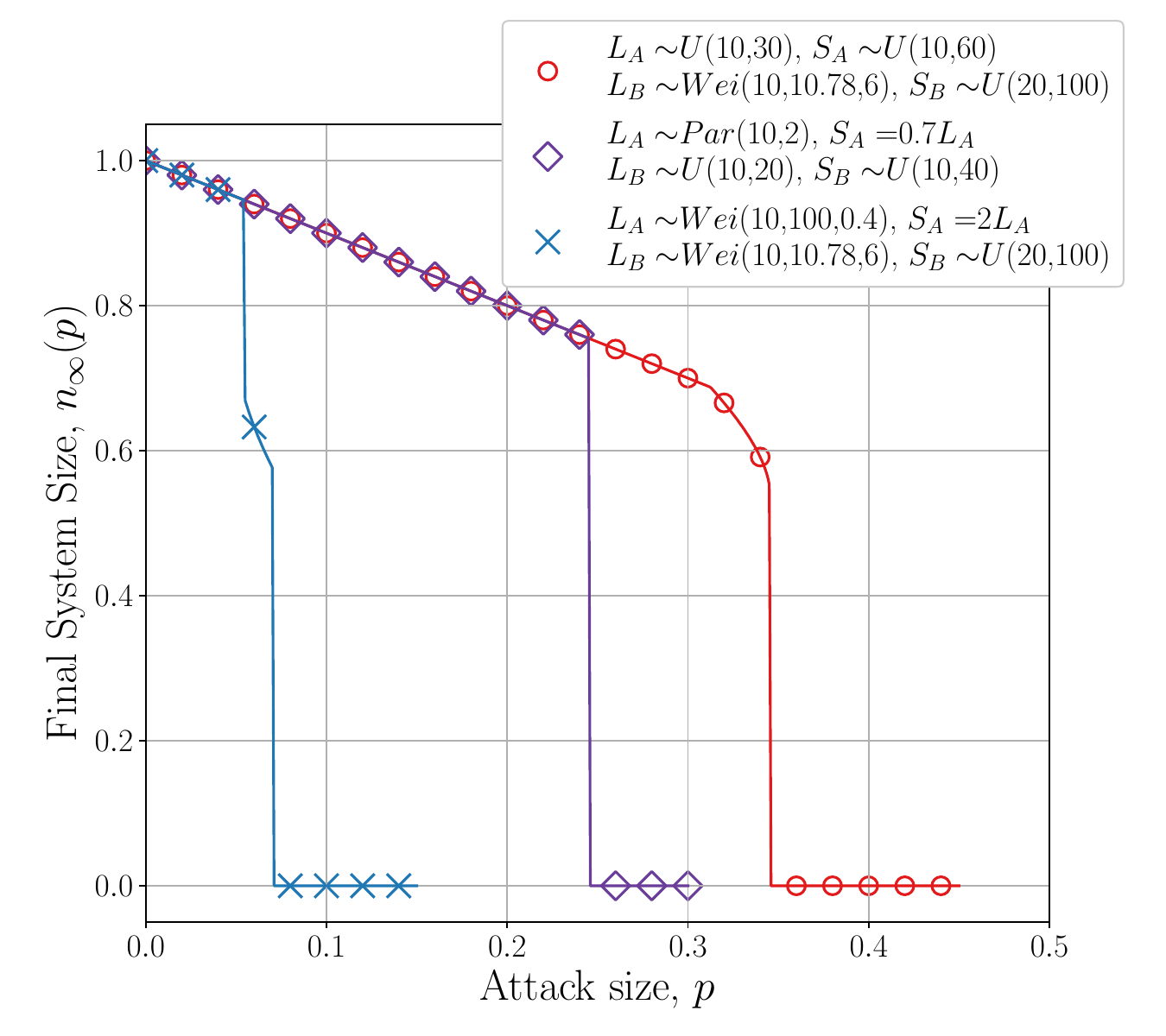}  
    \caption{Final system size for different $L_A,L_B,S_A,S_B$ distributions under random attack $p$. Analytical results are obtained by (\ref{eq_main:solution set}) are shown in solid lines where the averaged results from 100 independent simulations are shown with different markers. We see that theoretical results perfectly match with the simulation averages.}
    \label{fig_main:matching figure}
\end{figure}

Figure \ref{fig_main:matching figure} shows the final system size in response to the random failure of a fraction $p$ of nodes. 
Each color in the legend corresponds to a different configuration of $L_A, L_B, S_A$, and $S_B$ distributions, indicating the family of distributions and the parameters selected. 
The parameters for the uniform distribution are presented as $U(U_{\text{min}}, U_{\text{max}})$, where $U_{\text{min}}$ is the minimum value and $U_{\text{max}}$ is the maximum value. 
For the Weibull distribution, they are denoted as $Wei(W_{\text{min}}, \lambda, k)$, where $W_{\text{min}}$ is the minimum value, $\lambda$ is the scale parameter, and $k$ is the shape parameter. 
For the Pareto distribution, they are given as $Par(P_{\text{min}}, b)$, where $P_{\text{min}}$ is the minimum value and $b$ is the shape parameter.
For all configurations $\beta_A = \beta_B = 0.1$. 
The markers (circle, diamond, and cross) indicate the average results from 100 simulation runs with $N=5.10^5$, while the lines are obtained using (\ref{eq_main:solution set}).

First, we observe that the selected families of distributions exhibit diverse types of phase transitions: a first-order transition for the purple configuration, a second-order followed by a first-order transition for the red configuration, and a first-order transition followed by a second-order and another first-order transition for the blue-cross configuration. 
This demonstrates that the rich behavior observed in single-layer flow networks, as noted in \cite{single_flow_optimizing}, generalizes to the multiplex version of the problem as well.

Overall, the calculated theoretical values align well with the simulated ones, even in highly nonlinear cases such as the configuration denoted with blue-cross, where we observe two modes of phase transitions.  
It is also worth noting that, although we used $N = 5 \times 10^5$, for simple family of distributions, a good match can be achieved even with smaller system sizes, such as $N = 10^4$, for simple distribution families.  
This, in a sense, indicates that even though the calculations reflect the final surviving fraction in a mean-field sense, the inferences still hold for relatively small networks.  
This observation increases the potential applicability of the results to real-world networks.  
\begin{figure}[!h]
    \centering
    \includegraphics[width=0.98\linewidth]{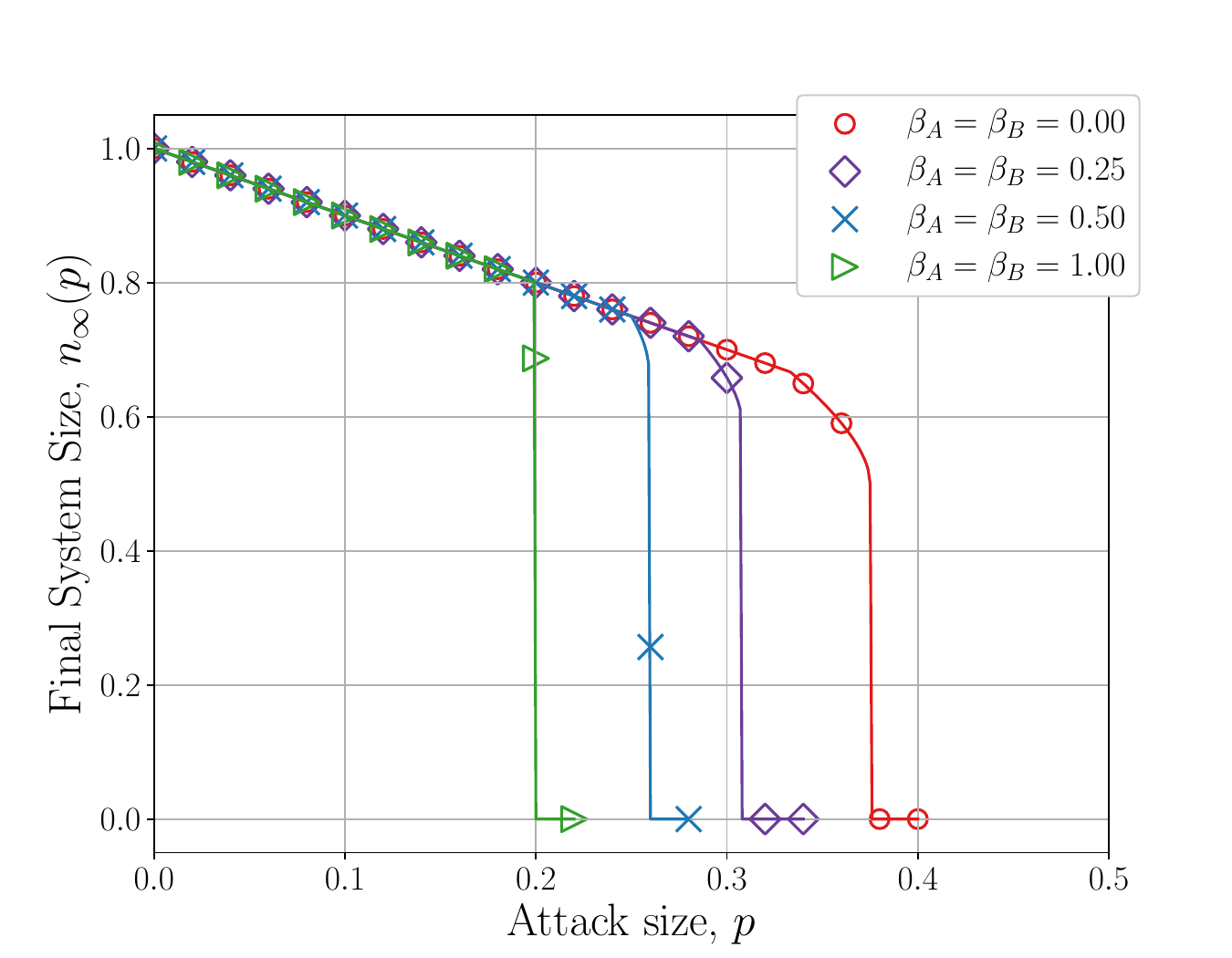}
    \caption{Final system size for $L_A \sim U(10,30)$, $L_B\sim Wei(10,10.78,6)$, $ S_A \sim U(10,60)$, $S_B \sim U(20,100)$ distributions under random attack $p$. Analytical results are obtained by (\ref{eq_main:solution set}) are shown in solid lines where the averaged results from 100 independent simulations with $N=10^5$ are shown with different markers. We see that theoretical results perfectly match with the simulation averages.}
    \label{fig_main:different beta}
\end{figure}

Figure \ref{fig_main:different beta} illustrates the effect of changing \textit{cross-layer influence factors} for the same initial load/free space distributions.  
Notably, the case where $\beta_A = \beta_B = 0$ corresponds to \textit{layer-independent overload}, confirming that our calculations for \textit{layer-influenced overload} are also applicable to this scenario.  
As indicated by lower critical attack sizes, networks with higher $\beta$ demonstrate reduced robustness due to the additional load transferred between layers.  
Furthermore, we observe that increasing $\beta$ not only impacts robustness but also alters the nature of the phase transition.  
Specifically, when $\beta = 1$, the second-order phase transition disappears entirely, leaving only a first-order transition.  
These findings emphasize the effect of inter-layer influence on both the robustness and dynamic behavior of multiplex networks under random attacks.  

\subsection{Numerical Validation of Optimal $L$,$S$ Distributions}
\begin{figure*}[!t]
    \centering
    \subfigure[Loads follow Weibull and Pareto Distribution]{\includegraphics[width = 0.48\linewidth]{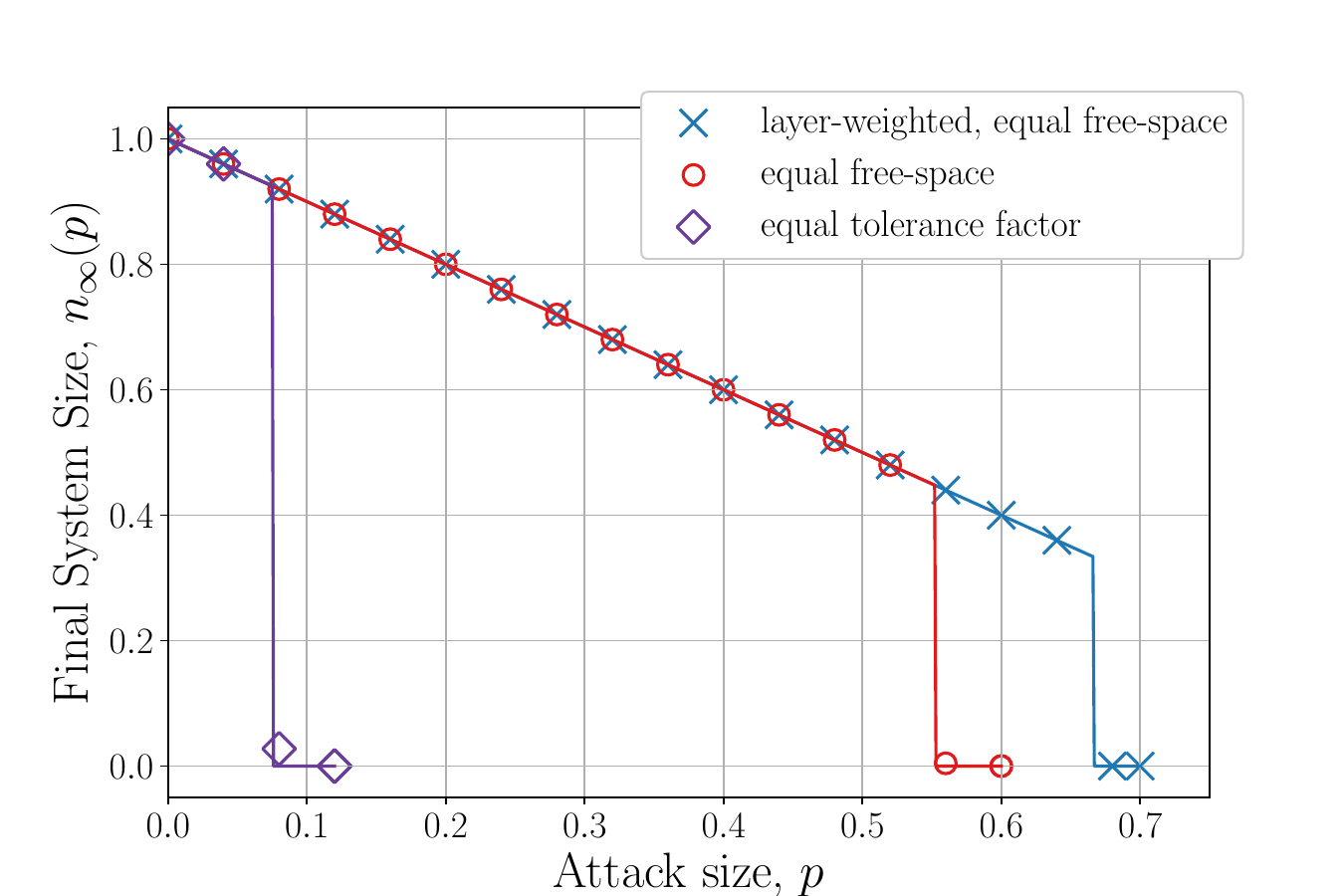}}
    \subfigure[Loads follow Pareto and Uniform Distribution]{\includegraphics[width = 0.48\linewidth]{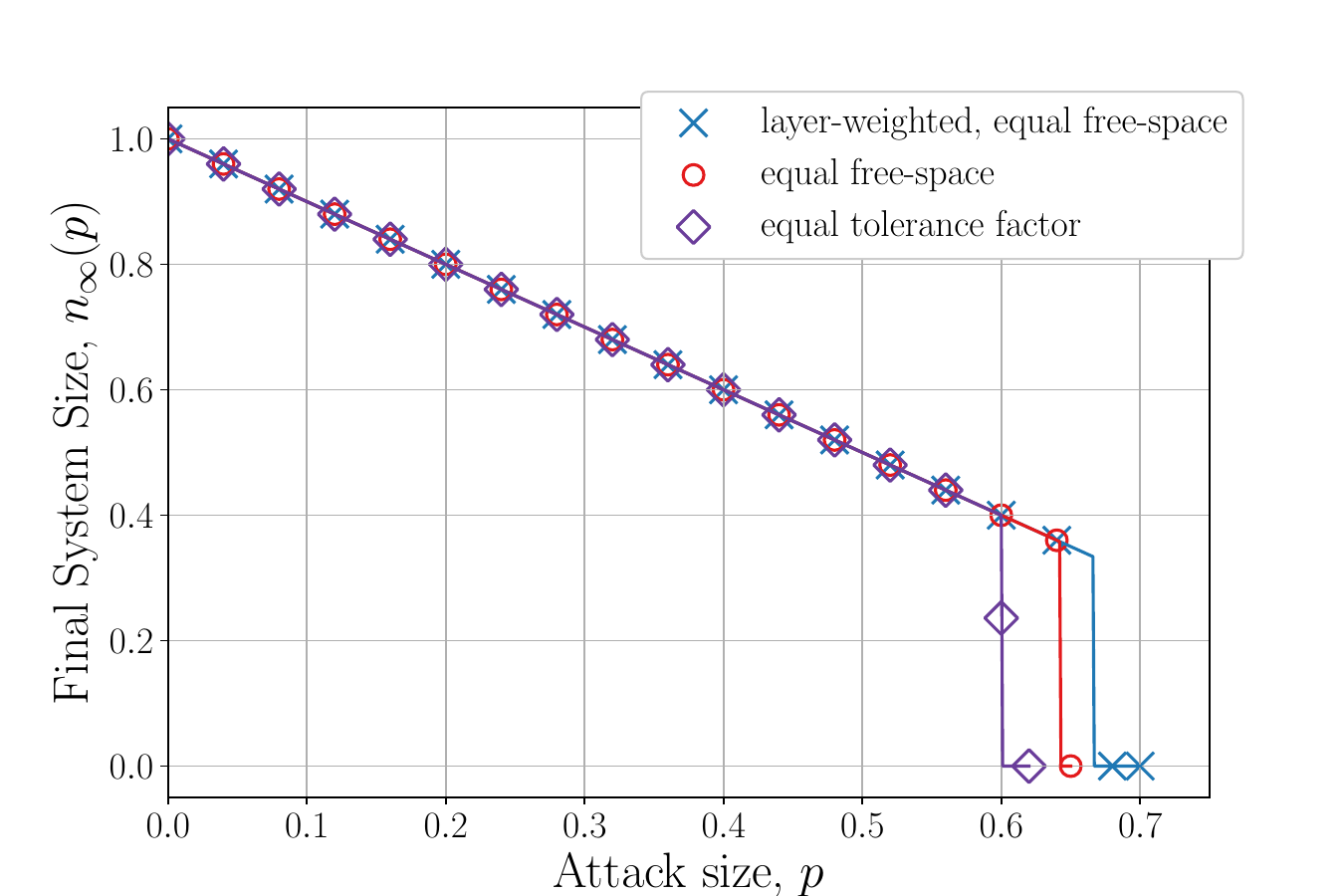}}\\
    \subfigure[Loads follow Uniform and Weibull Distribution]{\includegraphics[width = 0.48\linewidth]{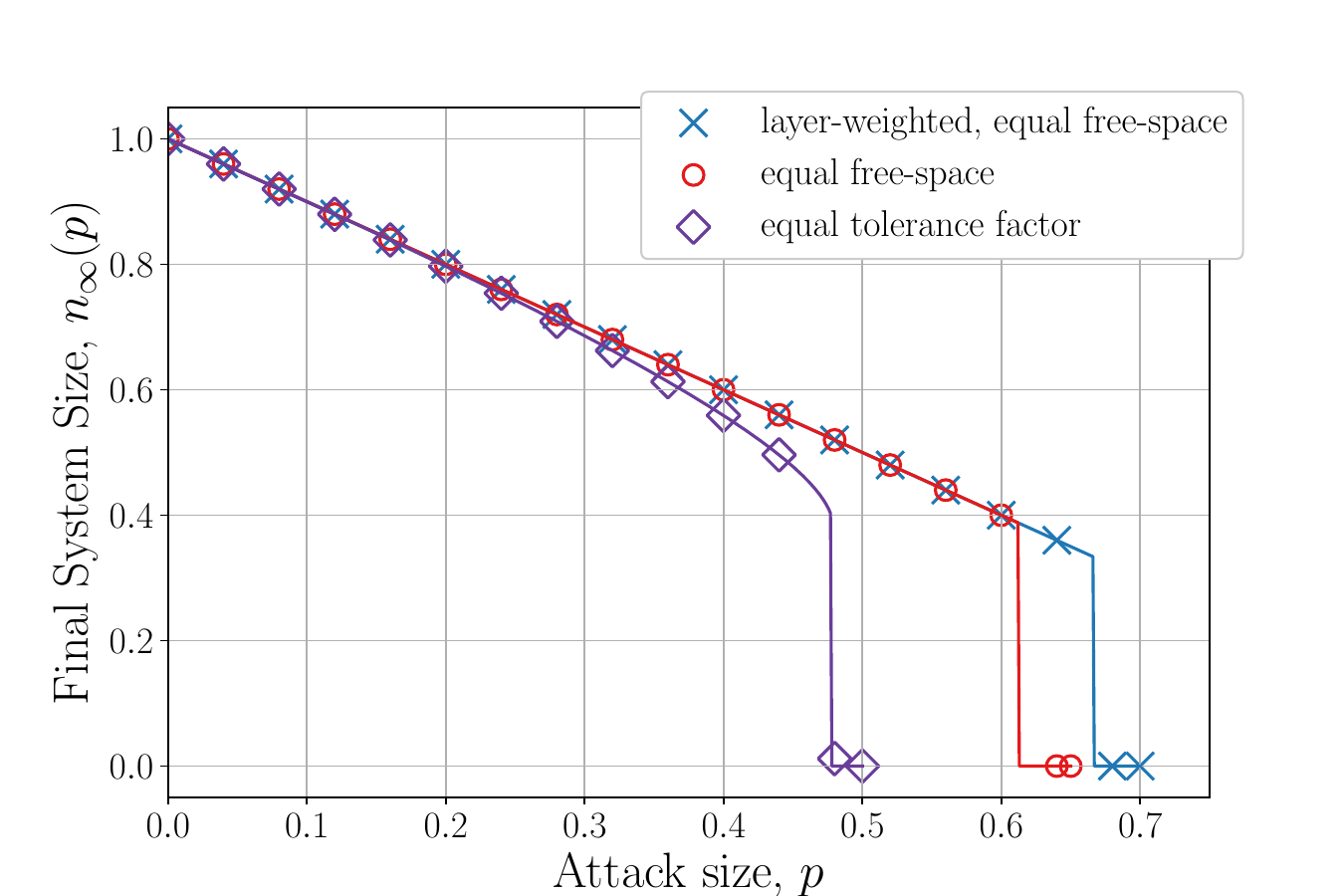}} 
    \subfigure[Comparison of optimal strategy for different configurations]{\includegraphics[width = 0.48\linewidth]{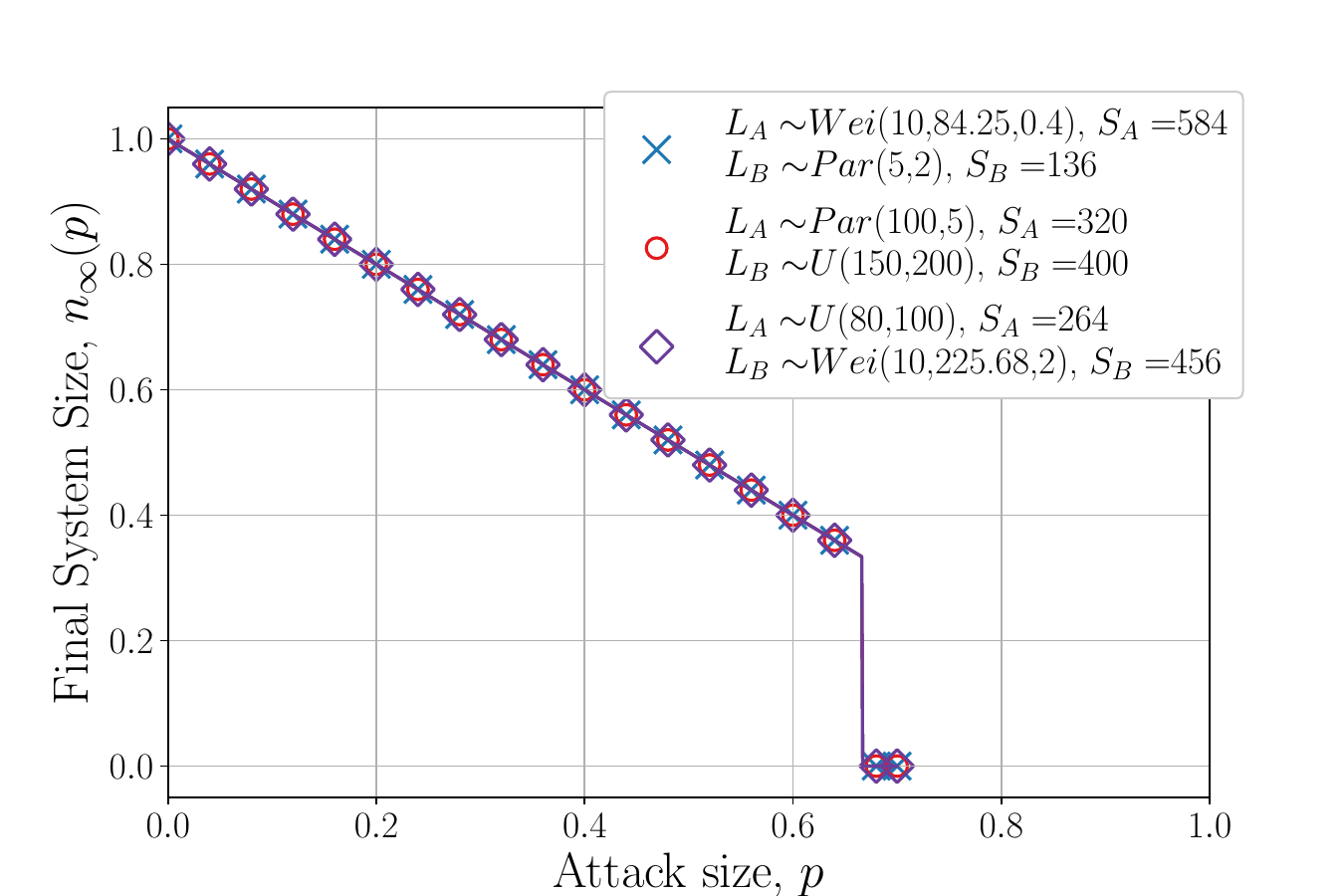}} 
    \caption{Final system sizes under different free space allocation strategies for three configurations: (a) $L_A \sim Wei(10,84.25, 0.4),L_B \sim Par(5,2)$, (b) $L_A \sim  Par(100,5), L_B \sim  U(150,200)$, (c) $L_A \sim U(80,100), L_B \sim Wei(10,225.68,2)$. For all configurations we have  $\mathbb{E}[L_A] + \mathbb{E}[L_B] = 300$, $\mathbb{E}[S_A] + \mathbb{E}[S_B] = 720$, and $\beta = 0.2$. For \textit{equal free space} strategy both layers receive $300$ free space and that space is distributed equally among the nodes in the layer. For \textit{equal tolerance factor}, $\alpha = 2.4$ to satisfy the numerical values of total expected load and the total expected free space. For \textit{layer-weighted, equal free space} strategy, the total free space is allocated each layer proportional to the expected loads adjusted with the \textit{cross-layer influence factors} ($\beta_A,\beta_B$), then it is distributed evenly among nodes in the layer. This results in the following free space value for each configurations: (a) $S_A = 584,S_B=136$, (b) $S_A = 320, S_B =400$, (c) $S_A = 264,S_B=456$. Subplot (d) presents the result of the optimal strategy for each configuration, which results in perfect overlap of the three configuration as they have the same total expected load and total expected free space values.}
    \label{fig:optimality_figure}
\end{figure*}

The second set of numerical simulations was conducted to validate our results on the optimality of the proposed strategy. 
For a single-layer flow network, \cite{single_flow_optimizing} discusses that the \textit{equal free space} scheme, where the total free space is distributed evenly among the nodes, outperforms the commonly used alternative \textit{equal tolerance factor} scheme \cite{alpha1,flow_2,hong_suppressing_2016,alpha2,pei_et.al.,artime_abrupt_2020,Wang_Jin_Zhao_2021}, in which the free space of a node is allocated proportionally to its load using a multiplicative factor $\alpha$ for all nodes. 
In the case of \textit{multiplex flow-network model}, an additional question of interest is how to allocate the total free space between the layers first and then determine how to distribute it among the nodes within each layer. 
To address this, we experimented with three different strategies:
\begin{itemize} 
    \item \textit{Layer-weighted, equal free space}: The total free space is allocated between Network $A$ and Network $B$ in proportion to their expected loads adjusted with \textit{cross-layer influence factors}, $\mathbb{E}[L_A]+\beta_B \mathbb{E}[L_B] $ and $\mathbb{E}[L_B]+\beta_A \mathbb{E}[L_A]$, respectively. Then, the free space within each layer is distributed evenly across the nodes in that layer. Notice that this corresponds to the optimality condition suggested by \textit{Theorem}~\ref{thm: Optimality}.
    \item \textit{Equal free space}: The total free space is allocated equally between the two layers. Subsequently, the free space within each layer is distributed evenly across the nodes in that layer. 
    \item \textit{Equal tolerance factor}: The free space for a node is allocated proportionally to its load. Note that this approach naturally results in a layer-weighted allocation across layers. 
\end{itemize}

Figure  \ref{fig:optimality_figure} presents the final fraction of surviving nodes for three different configurations using the three proposed strategies. 
For the configurations, we selected loads from different distribution families while keeping the $\beta_A =\beta_B = 0.2$, expected load $(\mathbb{E}[L_A] + \mathbb{E}[L_B] = 300)$ and free space $(\mathbb{E}[S_A] + \mathbb{E}[S_B] = 720)$ fixed across all configurations. 
The figure confirms that the \textit{layer-weighted, equal free space} strategy outperforms the other two strategies, validating our optimality results for a variety of different load/free space distribution families.

Figure \ref{fig:optimality_figure}(d) shows the results of the optimal strategy across the three configurations which are in perfect overlap. This supports our findings on the maximum critical attack size, as (\ref{eq_main:critical_attack_size_opt}) suggests $p_{opt}^* = 0.66$, marking the critical attack size in all configurations.
We also observe that the final system size $n_\infty(p)=1-p$ for all $p<p_{opt}$ indicating there are no additional failures after the initial attack. 
This indicates that performance of \textit{layer-weighted, equal free space} remains consistent across different load distribution families, whereas the performance of the other strategies varies significantly, with \textit{equal tolerance factor} performing substantially worse in many cases.

Overall, the experiments with the optimal strategy highlight several key strengths of the results.  
Under the \textit{layer-weighted, equal free space} configuration, the resulting final system size reaches its maximum, with no additional failures after the initial attack for any attack size. 
Moreover, the strategy does not depend on the shape of the initial load distribution or the individual initial loads of the nodes, rather it requires only knowledge of the expected loads.  
This strategy remains effective in engineered systems where complete information is unavailable, but some aggregate-level information is accessible.  
Additionally, the intuitive nature of \textit{layer-weighted, equal free space} strategy provides a high-level guideline for real-world systems, such as workload or financial networks, where detailed numerical analysis of the system may be infeasible.  

\section{Conclusion}
\label{sec:discussions and conc}
In this work, we proposed and analyzed the \textit{multiplex flow-network model} to study the robustness of flow networks against random failures in a multiplex setting.  
Our model captures a broad range of systems where nodes have distinct functionalities represented across multiple layers, with each functionality associated with a different type of flow, such as collaboration networks or distributed computing infrastructures.  
We provide recursive equations and their solutions, which enable the calculation of the final fraction of surviving nodes under random failures without requiring extensive simulations.  
Our findings on the optimal strategy reveal that the \textit{layer-weighted equal free space} strategy ensures maximum robustness in terms of the final system size for any attack size, while also offering an intuitive interpretation.  
Thus, the analysis presented provides valuable insights for both the design and management of such systems across various application domains.  

Regarding potential extensions, our study suggests several avenues for future research.  
First, we only considered two cases of overload conditions in this paper.
The same model can be utilized to explore more complex cases of failure conditions that allow nodes to perform their functions in layer $A$ but fail in layer $B$, while their failure still jointly determined by the loads in each layer.  
Second, while our analysis focused on system robustness under the \textit{global redistribution} rule, further research could investigate \textit{local redistribution} under different network topologies~\cite{pei_et.al.,Chen_Hu_Meng_Yu_2024,hong_suppressing_2016,Wang_Jin_Zhao_2021}.  
Additionally, the proposed framework remains applicable to more complex dependencies in \textit{layer-influenced overload}, beyond the linear increase in loads with respect to the loads in other networks considered here.
Third, building on prior work that examines the coupling of cyber-physical networks, where failures are influenced by both flow redistribution and structural connectivity \cite{zhang_robustness_2020}, the same idea can be expanded to the multiplex setting where where a multiplex flow network is coupled with a multiplex cyber network. 
Finally, future work could explore robustness strategies against \textit{targeted attacks}\cite{Wang_Jin_Zhao_2021,Zhou_Elmokashfi_2017,Zhang_Cheng_Zhao_Li_Lu_Wang_Xiao_2013,Ma_Xin_2024}, where the initially attacked nodes are selected strategically to maximize the impact of the attack, rather than the random attack scenario examined in this study.

\bibliographystyle{IEEEtran}
\bibliography{IEEEabrv,ref}



\appendices
\section{A proof of Theorem~\ref{thm:recursive equations}}
\label{sec:apx_final size}

Consider an initial attack that affects a fraction $p$ of nodes.  
Let the fraction of survived nodes at iteration $t$ be denoted as $n_t$, and let the extra load per surviving node after failures at iteration $t$ for layer $A$ and layer $B$ be denoted as $Q_{t,A}, Q_{t,B}$ respectively.  
For convenience of notation, we define the \textit{effective excess load} in each layer, denoted as $Q'_{t,i}$, as:
\begin{align*}
    Q'_{t,A}=Q_{t,A} + \beta_B Q_{t,B}, \\
    Q'_{t,B}=Q_{t,B} + \beta_A Q_{t,A}.
\end{align*}

Here, $\beta_A$ and $\beta_B$ are the \textit{cross-layer influence factors} representing the unit influence of the load in one layer on another, for layers $A$ and $B$, respectively.
It is important to note that, when analyzing failures, we must consider the \textit{effective excess loads} ($Q'_{A}, Q'_{B}$) rather than the actual excess loads ($Q_{A}, Q_{B}$).  
This adjustment captures the interdependent effects between layers, ensuring an accurate representation of load redistribution dynamics during successive failure iterations for \textit{layer-influenced overload} case with linear boundaries.

Then, for the initial conditions with the attack size of $p$, we have $n_0 = 1-p$.
Since the initially failed nodes are chosen randomly, we can calculate the resulting loads to redistribute as $Q_{0,A} = \frac{E[L_A]p}{(1-p)},Q_{0,B} = \frac{E[L_B]p}{(1-p)} $.
At the next iteration ($t = 1$), only nodes with more free space than $Q'_{0,A}$ and $Q'_{0,B}$ in their respective layers will survive.  
Then, the mean fraction of surviving nodes that fail at iteration $t = 1$ can be calculated as:
\begin{equation*}
    n_1 =  n_0\cdot \mathbb{P}[S_A > Q'_{A,0}, S_B>Q'_{B_0}].
\end{equation*}
Upon these failures at $t=1$, the loads of the failed nodes will be redistributed evenly across the remaining nodes. 
Next, we calculate $Q_{1,A}$ and $Q_{1,B}$ based on $n_1$. 
The total excess load at $t = 1$ consists of two components:  
(i) the total load of nodes that have already failed in the initial attack, and  
(ii) the total load of nodes that failed during the first iteration due to the effective excess loads $Q'_{0,A}$ and $Q'_{0,B}$.  
Focusing on layer $A$, we can express the total excess load at $t = 1$ as:
\begin{equation*}
    N\cdot n_1\cdot Q_{1,A} = N\cdot n_0\cdot Q_{0,A} + \mathbb{E}\left[ \sum_{x \notin \mathcal{I},x \in \mathcal{F_0}} L_{x,A} \right].
\end{equation*}
where $\mathcal{I}$ denoting the initially failed set of nodes, and $\mathcal{F_0}$ denoting the set of nodes satisfying the failure condition, i.e., $S_{x,A} < Q_{0,A} + \beta_B Q_{0,B}$ or $S_{x,B} < Q_{0,B} + \beta_A Q_{0,A}$.  

On the left-hand side, $N.n_1.Q_{1,A}$ represents the total excess load at iteration $t = 1$.  
The first term on the right-hand side, $N\cdot n_0\cdot Q_{A,0}$, accounts for the excess load carried over from iteration $t = 0$.  
The second term denotes the expected sum of loads from nodes that survived until the previous iteration but failed at iteration $t = 1$ due to the effective excess loads.  

By rearranging this expression, we can derive an equation for $Q_{1,A}$ in terms of the previous excess loads and the fraction of failed nodes:
\begin{multline}
         n_1\cdot Q_{1,A} =n_0\cdot Q_{0,A} \\ + \frac{1}{N} \mathbb{E}\left[\sum_{x \notin I} L_{x,A}\mathbbm{1}\left[
     \begin{array}{c}
           S_{x,A} < Q'_{0,A} \\
            \cup \\
        S_{x,B} < Q'_{0,B}
     \end{array}
       \right]  \right]
    \label{eq_apx:excess_load_calc_1}
\end{multline}
\begin{equation*}
    = (1-p)\frac{p\mathbb{E}[L_A]}{(1-p)} + \frac{1}{N} \mathbb{E}\left[\sum_{x \notin I} L_{x,A}(1 - \mathbbm{1}\left[
    \begin{array}{c}
       S_{x,A} > Q'_{0,A}  \\ S_{x,B} > Q'_{0,B}
    \end{array}
    \right])  \right]
\end{equation*}
\begin{equation*}
    = p \mathbb{E}[L_A] + (1-p) \mathbb{E}[L_A]-(1-p) \mathbb{E}\left[L_A\mathbbm{1}{\left[\begin{array}{c} S_A > Q'_{0,A}\\ S_B > Q'_{0,B} 
         \end{array}\right]} \right]
\end{equation*}
Therefore, we can compute the excess load per surviving node at iteration 1 as:
\begin{equation*}
    Q_{1,A} = \frac{\mathbb{E}[L_A] - (1-p)\mathbb{E}\left[L_A . \mathbbm{1}{\left[\begin{array}{c} S_A > Q'_{0,A}\\ S_B > Q'_{0,B} 
         \end{array}\right]}\right]}{n_{1}}
\end{equation*}
\begin{equation*}
    Q_{1,B} = \frac{\mathbb{E}[L_B] - (1-p)\mathbb{E}\left[L_B . \mathbbm{1}{\left[\begin{array}{c} S_A > Q'_{0,A}\\ S_B > Q'_{0,B} 
         \end{array}\right]}\right]}{n_{1}}
\end{equation*}
Now that we have calculated the excess load per surviving node in layers $A$ and B, we can proceed to $t=2$ and calculate the fraction of remaining nodes.  
From the previous iteration, we know that the remaining fraction was $n_1$.  
At this stage, nodes with free space larger than $Q'_{A,0}$ and $Q'_{B,0}$ but smaller than $Q'_{A,1}$ and $Q'_{B,1}$ will fail.  
Thus,
\begin{equation}
         n_{2} = n_1 \cdot  \mathbb{P}\left(\begin{array}{c|c} S_A > Q'_{A,1} & S_A > Q'_{A,0} \\ S_B > Q'_{B,1} & S_B > Q'_{B,0} \end{array}\right)
    \label{eq_apx:failed_nodes_at_2}
\end{equation}
Notice that (\ref{eq_apx:failed_nodes_at_2}) includes the $Q$ and $n$ values from previous iterations, and therefore, it can be calculated recursively.  
If we can express $Q_{A,2}$ and $Q_{B,2}$ in terms of previously calculated measures, we can establish a recursion for the final system size.  
Using the same logic as in (\ref{eq_apx:excess_load_calc_1}), we can write $Q_{A,2}$ as:
\begin{multline*}
n_2\cdot Q_{A,2}=n_0 \cdot Q_{A,0}\\
+ \frac{1}{N} E\left[\sum_{x \notin I} L_{x,A}\mathbbm{1}\left[
\begin{array}{c}
   S_{x,A} < Q'_{A,1}  \\ \cup \\S_{x,B} < Q'_{B,1}
\end{array}\right]  \right]
\end{multline*}
\begin{equation*}
    = p \mathbb{E}[L_A] + (1-p) \mathbb{E}[L_A]-(1-p) \mathbb{E}\left[L_A\mathbbm{1}{\left[\begin{array}{c} S_A > Q'_{1,A}\\ S_B > Q'_{1,B} 
         \end{array}\right]} \right]
\end{equation*}

Therefore, $Q_{A,2}$ and $Q_{B,2}$ can be calculated as:
\begin{equation}
    Q_{A,2} = \frac{\mathbb{E}[L_A] - (1-p)(L_A . \mathbbm{1}{\left[\begin{array}{c} S_A > Q'_{A,1}\\ S_B > Q'_{B,1} 
         \end{array}\right]})}{n_{2}}
\end{equation}
\begin{equation}
    Q_{B,2} = \frac{\mathbb{E}[L_B] - (1-p)(L_B . \mathbbm{1}{\left[\begin{array}{c} S_A > Q'_{A,1}\\ S_B > Q'_{B,1} 
         \end{array}\right]})}{n_{2}}
\end{equation}

Finally, with $n_0=1-p$, $Q_{-1,A}=Q_{-1,B}=0$, and $Q_{0,A}=\frac{p\mathbb{E}[L_A]}{1-p},Q_{0,B}=\frac{p\mathbb{E}[L_B]}{1-p}$, the surviving fraction of nodes at iteration $t=1,2,\ldots$ can be calculated using the following recursive equations: 
\begin{align}
         &n_{t} = n_{t-1} \cdot \mathbb{P}\left(\begin{array}{c|c} S_A > Q'_{t-1,A} & S_A > Q'_{t-2,A} \\ S_B > Q'_{t-1,B} & S_B > Q'_{t-2,B} \end{array}\right)
    \label{eq_apx:final_size_v1}\\
    &Q_{t,A} = \frac{\mathbb{E}[L_A] - (1-p)\mathbb{E}\left[L_A . \mathbbm{1}{\left[\begin{array}{c} S_A > Q'_{t-1,A}\\ S_B > Q'_{t-1,B} 
         \end{array}\right]}\right]}{n_{t}}
    \label{eq_apx:excess_load_A_v1}\\
    &Q_{t,B} = \frac{\mathbb{E}[L_B] - (1-p)\mathbb{E}\left[L_B . \mathbbm{1}{\left[\begin{array}{c} S_A > Q'_{t-1,A}\\ S_B > Q'_{t-1,B} 
         \end{array}\right]}\right]}{n_{t}}
    \label{eq_apx:excess_load_B_v1}
\end{align}

Given the initial distributions for $L_A$, $L_B$, $S_A$, and $S_B$, the final fraction of surviving nodes under a random attack of size $p$ can be calculated using (\ref{eq_apx:final_size_v1})-(\ref{eq_apx:excess_load_B_v1}), iteratively, until the stopping condition $n_{t+1} = n_{t}$ is met.  

(\ref{eq_apx:final_size_v1}) presents a unique structure which allows further simplification. 
Since $Q'_{t,A}$ and $Q'_{t,B}$ are monotonically increasing with $t$ (i.e., $Q'_{A,t+1} \geq Q'_{t,A}$ and $Q'_{B,t+1} \geq Q'_{t,B}$), repeated application of (\ref{eq_apx:final_size_v1}) yields:
\begin{align*}
        n_{t} &= n_{t-1}\frac{\mathbb{P}[S_A>Q'_{t-1,A},S_B>Q'_{t-1,B}]}{\mathbb{P}[S_A>Q'_{t-2,A},S_B>Q'_{t-2,B}]} \\
        n_{t-1} &= n_{t-2}\frac{\mathbb{P}[S_A>Q'_{t-2,A},S_B>Q'_{t-2,B}]}{\mathbb{P}[S_A>Q'_{A,t-3},S_B>Q'_{B,t-3}]}\\
        &\ .\\
        &\ .\\
        &\ .\\
        n_{2} &= n_1\frac{\mathbb{P}[S_A>Q'_{A,1},S_B>Q'_{B,1}]}{\mathbb{P}[S_A>Q'_{A,0},S_B>Q'_{B,0}]}\\
        n_{1} &= n_{0}\mathbb{P}[S_A>Q'_{A,0},S_B>Q'_{B,0}]
\end{align*}

Therefore, by multiplying both sides, we arrive at a simplified expression for $n_{t}$:
\begin{equation}
    n_{t} = n_0\cdot \mathbb{P}[S_A>Q'_{t-1,A},S_B>Q'_{t-1,B}].
    \label{eq_apx:final_size_v2}
\end{equation}

Substituting (\ref{eq_apx:final_size_v2}) into (\ref{eq_apx:excess_load_A_v1})-(\ref{eq_apx:excess_load_B_v1}), we obtain the finalized versions of the recursion formulas:
\begin{align}
    n_{t} &= n_0\cdot \mathbb{P}[S_A>Q'_{t-1,A},S_B>Q'_{t-1,B}]
    \label{eq_apx:final_size_N_v2}\\
Q_{t,A} &= \frac{\mathbb{E}[L_A] - (1-p)\mathbb{E}\left[L_A . \mathbbm{1}{\left[\begin{array}{c} S_A > Q'_{t-1,A}\\ S_B > Q'_{t-1,B} 
         \end{array}\right]}\right]}{n_0 \cdot \mathbb{P}[S_A>Q'_{t-1,A},S_B>Q'_{t-1,B}]}
    \label{eq_apx:excess_load_A_v2} \\ 
Q_{t,B} &= \frac{\mathbb{E}[L_B] - (1-p)\mathbb{E}\left[L_B . \mathbbm{1}{\left[\begin{array}{c} S_A > Q'_{t-1,A}\\ S_B > Q'_{t-1,B} 
         \end{array}\right]}\right]}{(n_0\cdot \mathbb{P}[S_A>Q'_{t-1,A},S_B>Q'_{t-1,B}]}
    \label{eq_apx:excess_load_B_v2} 
\end{align}
\hspace*{\fill}$\blacksquare$

\section{A proof of Theorem~\ref{thm:solution set}}
\label{sec:apx_fixed point solution}

Given (\ref{eq_apx:final_size_v1}), the stopping condition $n_{t+1}=n_t$ would imply:
\begin{equation}
\footnotesize
\mathbb{P}\left[\begin{array}{c|c} S_A > Q_{A,t+1} + \beta_B Q_{B,t+1} & S_A > Q_{t,A} + \beta_B Q_{t,B} \\ S_B > Q_{B,t+1} + \beta_A Q_{A,t+1} & S_B > Q_{t,B} + \beta_A Q_{t,A} \end{array}\right]=1
    \label{eq_apx:stopping cond}
\end{equation}

which happens either one of the cases apply as a solution:
\begin{enumerate}[(i)]
    \item $Q_{t,A} \geq Q_{A,t+1}$, $Q_{t,B}  \geq Q_{B,t+1}$
    \item $Q_{t,A} < Q_{A,t+1}$ and/or $Q_{t,B}<Q_{B,t+1}$ and (\ref{eq_apx:stopping cond})
\end{enumerate}

\begin{lemma}
    \label{lemma:Claim 1}
    For any solution of ($Q_{t,A},Q_{t,B}$) satisfying (ii), there is an equivalent solution in (i) yielding the same final fraction of surviving nodes. 
\end{lemma} 

\begin{IEEEproof}
    Consider a solution $(Q^*_{t,A}, Q^*_{t,B})$ satisfying condition (ii).  
    First, notice that the stopping condition, $n_t=n_{t+1}$ requires the final system sizes at iterations $t$ and $t+1$ to be identical.  
    Given (\ref{eq_apx:final_size_N_v2}), this implies that:
    \begin{equation*}
    \footnotesize
            \mathbb{P}\left[ \begin{array}{c}
            S_A > Q^*_{t,A} + \beta_B Q^*_{t,B}
            \\
            S_B > Q^*_{t,B} + \beta_A Q^*_{t,A}
       \end{array}  \right] = 
       \mathbb{P}\left[ \begin{array}{c}
            S_A > Q^*_{A,t+1} + \beta_B Q^*_{B,t+1} \\
            S_B > Q^*_{B,t+1} + \beta_A Q^*_{A,t+1}
       \end{array}  \right].
    \end{equation*}
    
    Using this equality, if we evaluate the excess load in the subsequent iteration $t+2$ using (\ref{eq_apx:excess_load_A_v2}), we obtain:
    \begin{flalign}
        Q^*_{A,t+2} &&
        \label{eq_apx:A_next_step_eq}
    \end{flalign}
    \begin{flalign*}
        &= \frac{\mathbb{E}[L_A] - (1-p)\mathbb{E}\left[L_A . \mathbbm{1}{\left[\begin{array}{c}
            S_A > Q^*_{A,t+1} + \beta_B Q^*_{B,t+1}
            \\
            S_B > Q^*_{B,t+1} + \beta_A Q^*_{A,t+1}
       \end{array} \right]}\right]}{(1-p).\mathbb{P}\left[\begin{array}{c}
            S_A > Q^*_{A,t+1} + \beta_B Q^*_{t,B}
            \\
            S_B > Q^*_{B,t+1} + \beta_A Q^*_{A,t+1}
       \end{array}\right]} &&
    \end{flalign*}
    \begin{flalign*}
    \footnotesize
        &= \frac{\mathbb{E}[L_A] - (1-p)\mathbb{E}\left[L_A . \mathbbm{1}{\left[\begin{array}{c}
            S_A > Q^*_{t,A} + \beta_B Q^*_{t,B}
            \\
            S_B > Q^*_{t,B} + \beta_A Q^*_{t,A}
       \end{array} \right]}\right]}{(1-p).\mathbb{P}\left[\begin{array}{c}
            S_A > Q^*_{t,A} + \beta_B Q^*_{t,B}
            \\
            S_B > Q^*_{t,B} + \beta_A Q^*_{t,A}
       \end{array}\right]}&&\\ 
       &= Q^*_{A,t+1} &&
    \end{flalign*}
    
    Similarly, for layer $B$, we can write:
    \begin{equation}
        Q^*_{B,t+2} = Q^*_{B,t+1} 
        \label{eq_apx:B_next_step_eq}
    \end{equation}
    
    (\ref{eq_apx:A_next_step_eq}) and (\ref{eq_apx:B_next_step_eq}) imply that the solution $(Q^*_{A,t+1}, Q^*_{B,t+1})$ satisfies condition (i) with $n_t=n_{t+1}$.  
    Therefore, for any solution satisfying condition (ii), there exists an equivalent solution that satisfies condition (i) and results in the same final system size which completes the proof.
\end{IEEEproof}

\textit{Lemma}~\ref{lemma:Claim 1} indicates that when we are looking for the fixed point solutions satisfying $n_{t+1}=n_t$, it is enough to consider only condition (i).
In the trivial case of complete failure, $Q_{t-1,A} = Q_{t-1,B} = \infty$ will result in $Q_{t,A} = Q_{t,B} = \infty$. Conversely, in the case of no initial attack, $Q_{0,A} = Q_{0,B} = 0$ will produce $Q_{1,A} = Q_{1,B} = 0$, leading to no failure at all.

Beyond these trivial cases, the system may exhibit no further failures after the initial attack if $Q_{0,A}=\frac{p\mathbb{E}[L_A]}{1-p}$ and $Q_{0,B}=\frac{p\mathbb{E}[L_B]}{1-p}$ satisfy the condition (i). 
Alternatively, the system may undergo several iterations following the initial attack, during which additional failures occur due to cascading effects. 
Ultimately, the system will converge to $Q_{t,A}$ and $Q_{t,B}$ values that satisfy the stopping condition (i). 
Thus, for a given system with defined $L_A, L_B, S_A, S_B$ distributions and a specified attack size, we define the set of points $(x_A,x_B)\in\mathcal{P}$ as \textit{stable points}, corresponding to $(Q_{t,A},Q_{t,B})$ values, if the excess load values $x_A$ and $x_B$ satisfy the condition (i).

\begin{lemma}
    \label{lemma:Claim 2}
    Let $\mathcal{P}$ be the set of \textit{stable points} for a given system and initial attack size $p$. 
    Let $ \mathbf{P}_1 = (a_1,b_1),\mathbf{P}_2= (a_2, b_2)$ be any two stable points, i.e. $\mathbf{P}_1,\mathbf{P}_2 \in \mathcal{P}$. 
    Then, the point $\mathbf{P}_3 = (\min (a_1,a_2),\min (b_1,b_2))$ is also a stable point, meaning $\mathbf{P}_3 \in \mathcal{P}$.
\end{lemma}
\begin{IEEEproof}
    If $a_1 \leq a_2$ and $b_1 \leq b_2$, then $\mathbf{P}_3 = (a_1, b_1) = \mathbf{P}_1 \in \mathcal{P}$, which is trivial.  
    Otherwise, if $a_1 < a_2$ and $b_1 > b_2$, we must show that $\mathbf{P}_3 = (a_1, b_2) \in \mathcal{P}$.  
    Since $\mathbf{P}_1$ is a stable point, it satisfies condition (i), implying:
    \begin{equation*}
        a_1 \geq \frac{\mathbb{E}[L_A] - (1-p)\mathbb{E}\left[L_A . \mathbbm{1}{\left[\begin{array}{c} S_A > a_1 + \beta_B b_1\\ S_B > b_1 +\beta_A a_1 
             \end{array}\right]}\right]}{(1-p) \mathbb{P}\left[\begin{array}{c} S_A > a_1 + \beta_B b_1\\ S_B > b_1 +\beta_A a_1 
             \end{array}\right] }
        \label{eq:corner_stable_proof_a}
    \end{equation*}
    
    Notice that, since we have $b_1 > b_2$, we can write:
    \begin{equation*}
        \mathbb{P}\left[\begin{array}{c} S_A > a_1 + \beta_B b_2\\ S_B > b_2 +\beta_A a_1 
             \end{array}\right] \geq \mathbb{P}\left[\begin{array}{c} S_A > a_1 + \beta_B b_1\\ S_B > b_1 +\beta_A a_1 
             \end{array}\right]
    \end{equation*}
    and 
    \begin{equation*}
    \footnotesize
        \mathbb{E}\left[L_A . \mathbbm{1}{\left[\begin{array}{c} S_A > a_1 + \beta_B b_2\\ S_B > b_2 +\beta_A a_1 
             \end{array}\right]}\right] \geq 
             \mathbb{E}\left[L_A . \mathbbm{1}{\left[\begin{array}{c} S_A > a_1 + \beta_B b_1\\ S_B > b_1 +\beta_A a_1 
             \end{array}\right]}\right]
    \end{equation*}
    
    Therefore, we can also conclude that:
    \begin{equation*}
        a_1 \geq \frac{\mathbb{E}[L_A] - (1-p)\mathbb{E}\left[L_A . \mathbbm{1}{\left[\begin{array}{c} S_A > a_1 + \beta_B b_2\\ S_B > b_2 +\beta_A a_1 
             \end{array}\right]}\right]}{(1-p) \mathbb{P}\left[\begin{array}{c} S_A > a_1 + \beta_B b_2\\ S_B > b_2 +\beta_A a_1 
             \end{array}\right] }
        \label{eq:corner_stable_proof_a_2}
    \end{equation*}
    
    Similarly, given that $\mathbf{P}_2$ is a stable point and $a_1<a_2$, we can write:
    \begin{align*}
        b_2 &\geq \frac{\mathbb{E}[L_B] - (1-p)\mathbb{E}\left[L_B . \mathbbm{1}{\left[\begin{array}{c} S_A > a_2 + \beta_B b_2\\ S_B > b_2 +\beta_A a_2 
             \end{array}\right]}\right]}{(1-p) \mathbb{P}\left[\begin{array}{c} S_A > a_2 + \beta_B b_2\\ S_B > b_2 +\beta_A a_2 
             \end{array}\right] }\\
        &\geq \frac{\mathbb{E}[L_B] - (1-p)\mathbb{E}\left[L_B . \mathbbm{1}{\left[\begin{array}{c} S_A > a_1 + \beta_B b_2\\ S_B > b_2 +\beta_A a_1 
             \end{array}\right]}\right]}{(1-p) \mathbb{P}\left[\begin{array}{c} S_A > a_1 + \beta_B b_2\\ S_B > b_2 +\beta_A a_1 
             \end{array}\right] }
    \end{align*}
    Therefore, the point $\mathbf{P}_3 = (a_1,b_2)$ also satisfies (i), indicating $\mathbf{P}_3 \in \mathcal{P}$. 
\end{IEEEproof}

\textit{Lemma}~\ref{lemma:Claim 2} demonstrates that among the stable points, there exists one that is element-wise minimum. 
This ensures that when searching for the solution with the maximum final size, or equivalently the minimum excess load, we can identify a stable point with corresponding to the smallest $Q_A$ and $Q_B$ values.

In sum, \textit{Lemma}~\ref{lemma:Claim 1} indicates that to search for \textit{stable points}, it is enough to consider the set:
\begin{equation*}
\mathcal{P}:\{x,y: x\geq g(x,y),y\geq h(x,y)\}
\end{equation*}
where
\begin{equation*}
g(x,y) = \frac{\mathbb{E}[L_A] - (1-p)\mathbb{E}\left[L_A . \mathbbm{1}{\left[\begin{array}{c} S_A > x + \beta_B y\\ S_B > y + \beta_A x 
         \end{array}\right]}\right]}{(1-p).\mathbb{P}[S_A>x + \beta_B y,S_B>y + \beta_A x]}
\end{equation*}
\begin{equation*}
h(x,y) = \frac{\mathbb{E}[L_B] - (1-p)\mathbb{E}\left[L_B . \mathbbm{1}{\left[\begin{array}{c} S_A > x+ \beta_B y\\ S_B > y + \beta_A x 
         \end{array}\right]}\right]}{(1-p).\mathbb{P}[S_A>x+ \beta_B y,S_B>y+ \beta_A x]}
\end{equation*}

Then, to find the resulting final system size, one should search for the \textit{stable point} that results in the maximum final fraction of surviving nodes, i.e. the minimum excess loads per networks.
\textit{Lemma}~\ref{lemma:Claim 2} assures that, given the existence of \textit{stable point(s)}, there is a solution $(x^*,y^*)$ such that $x^*$ and $y^*$ are element-wise minimum among all \textit{stable points}, meaning that $x^*\leq x \text{ and } y^* \leq y \text{ for all } (x,y)\in \mathcal{P}$. 
Then, the final system size can be calculated by:
\begin{equation*}
    n_{\infty}(p) = (1-p).\mathbb{P}[S_A>x^*+\beta_B y^*,S_B>y^*+\beta_A x^*].
\end{equation*}\hspace*{\fill}$\blacksquare$

\section{A proof of Theorem~\ref{thm: Optimality}}
\label{sec:apx_optimality}
To prove this, we first derive an upper bound for $p^*$, the maximum critical attack size of a two-layer multiplex network with general $L$ and $S$ distributions, subject to the constraint $\mathbb{E}[S_A] + \mathbb{E}[S_B] = S_{\text{total}}$.  
We then demonstrate that the critical attack size of the configuration described in our claim ($p^*_\text{Dirac}$) achieves this upper bound.

To obtain an upper bound on $p^*$, we start by considering the critical attack sizes of individual networks $A$ and $B$ ($p_A$ and $p_B$, respectively).  
Here, $p_A$ denotes the upper bound for the attack size that results in complete failure in layer A, assuming no additional failures occur due to loads in layer B, and vice versa.  

Yingrui and Yagan~\cite{single_flow_optimizing} provided the upper bound for single-layer flow networks as $p = \frac{\mathbb{E}[S]}{\mathbb{E}[\text{Capacity}]}$.  
Applying the same reasoning to our initial configuration for \textit{layer-influenced overload}, we can write $p_A$ and $p_B$ as:
\begin{align}
    p_A&=\frac{\mathbb{E}[S_A]}{\mathbb{E}[S_A]+\mathbb{E}[L_A]+\beta_B\mathbb{E}[L_B]}
    \label{eq_apx:critical_single_layer_A}
    \\   p_B&=\frac{\mathbb{E}[S_B]}{\mathbb{E}[S_B]+\mathbb{E}[L_B]+\beta_A\mathbb{E}[L_A]}
    \label{eq_apx:critical_single_layer_B}
\end{align}
Since there is one-to-one dependence between functionalities of the nodes in each layer, if any layer experiences complete failure, the entire system collapses.  
Therefore, the multiplex network can be at most as resilient as its weakest layer, which implies:
\begin{equation}
    p^* \leq \min(p_A, p_B).
    \label{eq_apx:upper_bound}
\end{equation}

Given the constraint $\mathbb{E}[S_A] + \mathbb{E}[S_B] = S_{\text{total}}$, we can substitute $\mathbb{E}[S_A] = x$, where $x \in [0, S_{\text{total}}]$, and rewrite (\ref{eq_apx:critical_single_layer_A}) and (\ref{eq_apx:critical_single_layer_B}) as:
\begin{align}
    p_A&=\frac{x}{\mathbb{E}[S_A]+\mathbb{E}[L_A]+\beta_B\mathbb{E}[L_B]}
    \label{eq_apx:critical_single_layer_A_v2}
    \\   p_B&=\frac{S_{\text{total}}-x}{\mathbb{E}[S_B]+\mathbb{E}[L_B]+\beta_A\mathbb{E}[L_A]}
    \label{eq_apx:critical_single_layer_B_v2}
\end{align}
Let the solution of $p_A(x) = p_B(x)$ be denoted as $x^*$, with the corresponding function value given by $p_A(x^*) = p_B(x^*) = p^*_\text{opt}$.
One can verify that such a solution exists, where:
\[
x^* = \frac{S_{\text{total}} (\mathbb{E}[L_A] + \beta_B \mathbb{E}[L_B])}{(1 + \beta_A)\mathbb{E}[L_A] + (1 + \beta_B)\mathbb{E}[L_B]},
\]

and:

\begin{equation}
\footnotesize
 p^*_\text{opt} = \frac{\mathbb{E}[S_A] +\mathbb{E}[S_B]}{\mathbb{E}[S_A] +\mathbb{E}[S_B]+(1+\beta_A)\mathbb{E}[L_A] +(1+\beta_B)\mathbb{E}[L_B]}
 \label{eq:p_AB}
\end{equation}
By analyzing the derivatives of (\ref{eq_apx:critical_single_layer_A_v2}) and (\ref{eq_apx:critical_single_layer_B_v2}), we find that $\frac{dp_A(x)}{dx} > 0$, implying $p_A(x) < p^*_\text{opt}$ for $x \in [0, x^*)$.  
Similarly, since $\frac{dp_B(x)}{dx} < 0$, we have $p_B(x) < p^*_\text{opt}$ for $x \in (x^*, S_{\text{total}}]$.  
Combining these results, we conclude:
\[
p^* \leq \min(p_A, p_B) \leq p^*_\text{opt}, \quad \text{for any } x \in [0, S_{\text{total}}].
\]
Substituting (\ref{eq:p_AB}) into these inequalities, we derive an upper bound for $p_\text{Dirac}$:
\[
p^*_{Dirac} \leq \frac{\mathbb{E}[S_A] +\mathbb{E}[S_B]}{\mathbb{E}[S_A] +\mathbb{E}[S_B]+(1+\beta_A)\mathbb{E}[L_A] +(1+\beta_B)\mathbb{E}[L_B]}
\]

Finally, we demonstrate that $p^*_\text{Dirac}$ achieves this upper bound.  
Consider the initial attack size:
\[
p = \left[\frac{\mathbb{E}[S_A] + \mathbb{E}[S_B]}{\mathbb{E}[S_A] + \mathbb{E}[S_B] + (1 + \beta_A)\mathbb{E}[L_A] + (1 + \beta_B)\mathbb{E}[L_B]}\right]^-.
\]

Using (\ref{eq_apx:final_size_v1})-(\ref{eq_apx:excess_load_B_v1}), we calculate:
\begin{align*}
    Q_{A,0}&=\frac{p}{1-p}\mathbb{E}[L_A] \\ &= \left[ \frac{S_{\text{total}}}{(1 + \beta_A)\mathbb{E}[L_A]+(1 + \beta_B)\mathbb{E}[L_B]} \right]^- \mathbb{E}[L_A],\\
    Q_{B,0}&=\frac{p}{1-p}\mathbb{E}[L_B] \\
    &= \left[ \frac{S_{\text{total}}}{(1 + \beta_A)\mathbb{E}[L_A]+(1 + \beta_B)\mathbb{E}[L_B]} \right]^- \mathbb{E}[L_B].
\end{align*}

Then we can calculate the effective excess loads in first iteration using $Q'_{A,0}=Q_{A,0}+\beta_BQ_{B,0}$ and $Q'_{B,0}=Q_{B,0}+\beta_AQ_{A,0}$:
\begin{equation*}
    Q'_{A,0}=\left[S_{\text{total}}\frac{\mathbb{E}[L_A]+\beta_B \mathbb{E}[L_B]}{(1+\beta_A)\mathbb{E}[L_A]+(1+\beta_B)\mathbb{E}[L_B]}\right]^-
\end{equation*}
\begin{equation*}
    Q'_{B,0}=\left[S_{\text{total}}\frac{\mathbb{E}[L_B]+\beta_A \mathbb{E}[L_A]}{(1+\beta_A)\mathbb{E}[L_A]+(1+\beta_B)\mathbb{E}[L_B]}\right]^-
\end{equation*}

Noting that $\mathbb{P_{S_A}} = \delta(x-S_{\text{total}}\frac{\mathbb{E}[L_A]+\beta_B \mathbb{E}[L_B]}{(1+\beta_A)\mathbb{E}[L_A]+(1+\beta_B)\mathbb{E}[L_B]})$ and $\mathbb{P_{S_B}} = \delta(x-S_{\text{total}}\frac{\mathbb{E}[L_B]+\beta_A \mathbb{E}[L_A]}{(1+\beta_A)\mathbb{E}[L_A]+(1+\beta_B)\mathbb{E}[L_B]})$ we have:

\begin{equation*}
    n_{1}= (1-p) \mathbb{P}\left[\begin{array}{c}
         S_A>Q'_{A,0}  \\
         S_B>Q'_{B,0}
    \end{array}\right] =1-p =n_0.
\end{equation*}

This implies the configuration survives the attack $p$ without any subsequent failures. Then the critical attack size of the configuration $p^*_\text{Dirac}=p^*_\text{opt}$ in (\ref{eq:p_AB}), meaning that 
\begin{multline*}
    p^*_\text{Dirac}= \\\left[ \frac{\mathbb{E}[S_A] +\mathbb{E}[S_B]}{\mathbb{E}[S_A] +\mathbb{E}[S_B]+(1+\beta_A)\mathbb{E}[L_A] +(1+\beta_B)\mathbb{E}[L_B]} \right].
    \label{eq:critical_attack_size_opt}
\end{multline*}
The calculation also indicates that the final system size $1-p$.
The same calculation will yield $n_1(p)=1-p$ for any attack size $p<p^*_\text{opt}$ indicating that the final system size of the system with this configuration can be expressed as:
\begin{equation*}
        n_{\infty,\text{Dirac}}=
        \begin{cases}
            1-p,&  p\leq p^*_\text{opt}\\
            0,              & p>p^*_\text{opt}
        \end{cases}
    \end{equation*}
\vspace{11pt}
\hspace*{\fill}$\blacksquare$

\begin{thm}
    \label{thm: Optimality_v2}
  {\sl   Consider the \textit{multiplex flow network model} described in \textit{Theorem}~\ref{thm:recursive equations}.  
    Let $\mathbb{E}[S_A]=\mu_A,\ \mathbb{E}[S_B] = \mu_B$ where $\mu_A$, $\mu_B$ are constants.
    \begin{itemize}
    \item[i)] Under these constraints, the critical attack size $p^*$ is upper-bounded as  $p^* \leq p^*_\text{opt}$, where
    \begin{align}
    & p_{opt}^* = \min (p_A,p_B)\\
    p_A&=\frac{\mu_A}{\mu_A+\mathbb{E}[L_A]+\beta_B\mathbb{E}[L_B]}
    \\   p_B&=\frac{\mu_B}{\mu_B+\mathbb{E}[L_B]+\beta_A\mathbb{E}[L_A]}    
    \end{align}
    \item[ii)]
    The optimal critical attack size,   $p^*_\text{opt}$,
 can be achieved by choosing the free space distributions as
    \begin{align*}
        p_{S_A} &= \delta\left(x-\mu_A\right)\\
        p_{S_B} &= \delta\left(x-\mu_B\right)
    \end{align*}   
    irrespective of how initial loads are distributed. 
   \item[iii)] The free space distributions given above maximizes the final system size, $n_{\infty}(p)$ for all $p$ in $(0,1)$. Namely, with $n_{\infty,\text{Dirac}}$ denoting the final system size under this configuration, we have:
    \begin{equation*}
        n_{\infty,\text{Dirac}}= 
        \begin{cases}
            1-p,&  p\leq p^*_\text{opt}\\
            0,              & p>p^*_\text{opt}
        \end{cases}
    \end{equation*}
    \end{itemize}
}
\end{thm}

\begin{IEEEproof}
The proof follows the same steps as the proof of Theorem~\ref{thm: Optimality}.  
Again, we define $p_A$, $p_B$ as the critical attack sizes of the individual layers, assuming no interdependence between them.  
Noting that \( \mathbb{E}[S_A]=\mu_A,\ \mathbb{E}[S_B] = \mu_B \), equations (\ref{eq_apx:critical_single_layer_A}) and (\ref{eq_apx:critical_single_layer_B}) yield:

    \begin{align}
    p_A&=\frac{\mu_A}{\mu_A+\mathbb{E}[L_A]+\beta_B\mathbb{E}[L_B]}
    \label{eq_apx:critical_single_layer_A_opt_2}
    \\   p_B&=\frac{\mu_B}{\mu_B+\mathbb{E}[L_B]+\beta_A\mathbb{E}[L_A]}
    \label{eq_apx:critical_single_layer_B_opt2}
\end{align}

The critical attack size (\( p^* \)) of the multiplex network with \textit{layer-influenced overload} is bounded by the critical attack size of the weakest layer.  
Thus, an upper bound \( p^*_\text{opt} \) for the critical attack size of the multiplex network is given by:
\begin{equation*}
    p^* \leq p^*_\text{opt} = \min(p_A, p_B).
\end{equation*}

Without loss of generality, let us assume that \( p_A \leq p_B \), meaning that \( p^*_\text{opt} = p_A \), which can be expressed as:
\begin{equation*}
    p^*_\text{opt} = \frac{\mu_A}{\mathbb{E}[S_A] + \mathbb{E}[L_A] + \beta_B \mathbb{E}[L_B]}.
\end{equation*}

Next, using the recursive equations, we show that the system with initial free space distributions \( p_{S_A} = \delta\left(x - \mu_A\right) \) and \( p_{S_B} = \delta\left(x - \mu_B\right) \) survives the attack size \( [p^*_\text{opt}]^- \).  

Consider the initial attack size:
\begin{equation*}
    p = \left[\frac{\mu_A}{\mu_A + \mathbb{E}[L_A] + \beta_B \mathbb{E}[L_B]}\right]^-.
\end{equation*}
Using equations (\ref{eq_apx:final_size_v1})-(\ref{eq_apx:excess_load_B_v1}), we compute:
\begin{align*}
    Q_{A,0}&=\frac{p}{1-p}\mathbb{E}[L_A] = \left[ \frac{\mu_A}{\mathbb{E}[L_A]+\beta_B\mathbb{E}[L_B]} \right]^- \mathbb{E}[L_A],\\
    Q_{B,0}&=\frac{p}{1-p}\mathbb{E}[L_B] = \left[ \frac{\mu_A}{\mathbb{E}[L_A]+ \beta_B\mathbb{E}[L_B]} \right]^- \mathbb{E}[L_B].
\end{align*}

Next, we calculate the effective excess loads in the first iteration using $Q'_{A,0}=Q_{A,0}+\beta_BQ_{B,0}$ and $Q'_{B,0}=Q_{B,0}+\beta_AQ_{A,0}$:

\begin{equation*}
    Q'_{A,0}=\left[\mu_A\right]^-
\end{equation*}
\begin{equation*}
    Q'_{B,0}=\left[\mu_A\frac{\mathbb{E}[L_B]+\beta_A \mathbb{E}[L_A]}{\mathbb{E}[L_A]+\beta_B\mathbb{E}[L_B]}\right]^-
\end{equation*}

Note that \( p_A \leq p_B \) implies:
\begin{align*}
    \frac{\mu_A}{\mu_A+\mathbb{E}[L_A]+\beta_B\mathbb{E}[L_B]} &\leq \frac{\mu_B}{\mu_B+\mathbb{E}[L_B]+\beta_A\mathbb{E}[L_A]}, \\
    \mu_A(\mathbb{E}[L_B]+\beta_A\mathbb{E}[L_A]) &\leq \mu_B(\mathbb{E}[L_A]+\beta_B\mathbb{E}[L_B]), \\
    \mu_A\frac{\mathbb{E}[L_B]+\beta_A\mathbb{E}[L_A]}{\mathbb{E}[L_A]+\beta_B\mathbb{E}[L_B]} &\leq \mu_B.
\end{align*}

Given the initial free space distributions \( p_{S_A} = \delta\left(x - \mu_A\right) \) and \( p_{S_B} = \delta\left(x - \mu_B\right) \), the surviving fraction of nodes can be calculated as:
\begin{equation*}
    n_{1}= (1-p) \mathbb{P}\left[\begin{array}{c}
         S_A > Q'_{A,0}  \\
         S_B > Q'_{B,0}
    \end{array}\right] = 1 - p = n_0.
\end{equation*}

This result implies that the system survives an attack of size \( [p^*_\text{opt}]^- \), which serves as an upper bound.  
Thus, the critical attack size of the configuration is given by \( p_\text{Dirac} = p^*_\text{opt} \).  
Furthermore, the final system size remains \( 1 - p \), which holds for any attack size \( p < p^*_\text{opt} \).  
Hence, the final system size in steady-state can be expressed as:
\begin{equation*}
        n_{\infty,\text{Dirac}}=
        \begin{cases}
            1 - p, &  p \leq p^*_\text{opt}, \\
            0,     & p > p^*_\text{opt}.
        \end{cases}
\end{equation*}

\vspace{11pt}

\end{IEEEproof}

\begin{IEEEbiography}[{\includegraphics[width=1in,height=1.25in,clip,keepaspectratio]{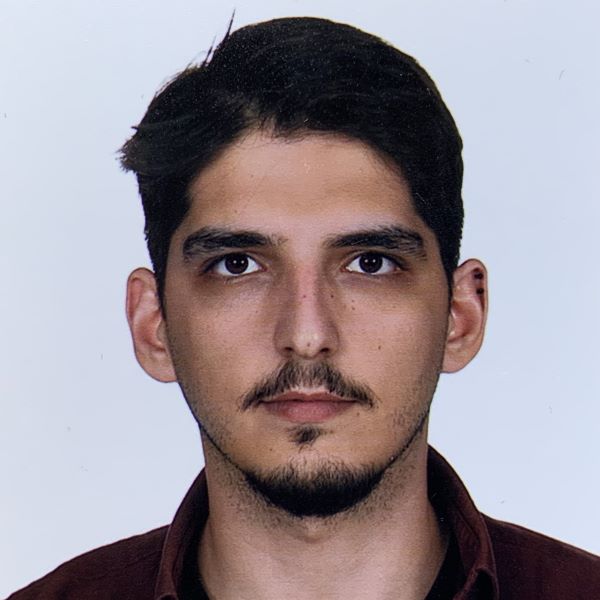}}]{Orkun İrsoy}
is a Ph.D. student at Electrical and Computer Engineering department in Carnegie Mellon University. He received his B.S. and M.S. degrees in Industrial Engineering from Boğaziçi University, Istanbul (Türkiye), in 2019 and 2022, respectively. His research broadly focuses on analyzing complex systems using network science, simulation, and data science, spanning a wide range of application areas, including engineering, health, and social systems."
\end{IEEEbiography}

\begin{IEEEbiography}[{\includegraphics[width=1in,height=1.25in,clip,keepaspectratio]{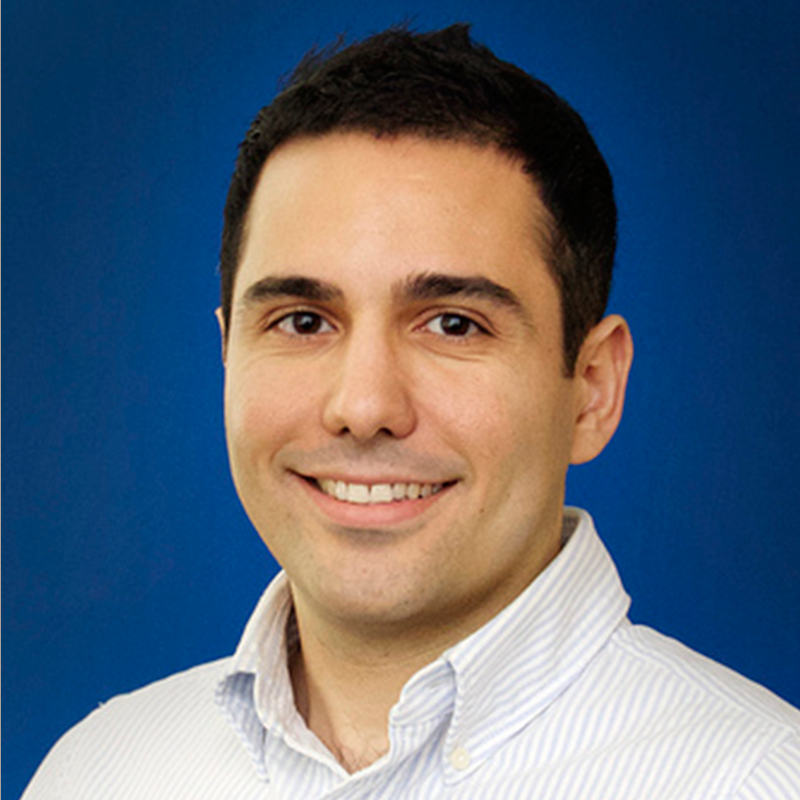}}]{Osman Yağan}
is a Research Professor of Electrical and Computer Engineering at Carnegie Mellon University. He received his Ph.D. degree in Electrical and Computer Engineering from the University of Maryland at College Park, MD in 2011, and his B.S. degree in Electrical and Electronics Engineering from the Middle East Technical University, Ankara (Turkey) in 2007.
 Dr. Yağan's research focuses on modeling, analysis, and performance optimization of computing systems, and uses tools from applied probability, network science, data science, and machine learning.  He is a recipient of a CIT Dean’s Early Career Fellowship, an IBM Academic Award, and best paper awards in ICC 2021, IPSN 2022, and ASONAM 2023.
\end{IEEEbiography}

\vfill

\end{document}